\documentclass[journal]{IEEEtran}
\usepackage{algorithm}
\usepackage[noend]{algpseudocode}

\usepackage{caption}
\DeclareCaptionFormat{cont}{#1 (continued)#2#3\par}
\usepackage{amsmath}
\usepackage{mathtools}
\usepackage{amsmath,amsfonts,amssymb}
\usepackage{booktabs}
\usepackage{mathrsfs}
\usepackage{makecell}
\usepackage{multirow}
\usepackage{soul,xcolor}
\usepackage[noadjust]{cite}
\DeclarePairedDelimiter\ceil{\lceil}{\rceil}

\usepackage[switch,columnwise]{lineno}
\usepackage{amssymb}
\usepackage{pifont}
\usepackage{xcolor,cite,etoolbox}
\makeatletter 
\pretocmd\@bibitem{\color{black}\csname keycolor#1\endcsname}{}{\fail}
\newcommand\citecolor[1]{\@namedef{keycolor#1}{\color{black}}}
\makeatother
\citecolor{maiwald2004comparison}
\citecolor{truong2019semi}
\citecolor{park2017modular}
\citecolor{shi2006sleep}
\citecolor{doyeon2021bmi}
\citecolor{muller2014minimally}
\citecolor{uehlin20190}
\citecolor{dilorenzo2019neural}
\citecolor{sharma2018acquisition}
\citecolor{musk2019integrated}
\citecolor{mandal202146}

\begin{document}
\setstcolor{black}
\title{Closed-Loop Neural Prostheses with On-Chip Intelligence: \textcolor{black}{A Review and A Low-Latency Machine Learning Model for Brain State Detection}}

\author{Bingzhao~Zhu,~\IEEEmembership{Student Member,~IEEE}, Uisub Shin, and Mahsa~Shoaran,~\IEEEmembership{Member,~IEEE}   
\vspace{-3mm}
\thanks{B. Zhu is with the School of Applied and Engineering Physics, Cornell University, Ithaca, NY, 14853 USA, and with  the  Institute of Electrical Engineering and Center for Neuroprosthetics, EPFL, 1202 Geneva, Switzerland  (e-mail: bz323@cornell.edu).}
\thanks{U. Shin is with the School of Electrical and Computer Engineering, Cornell University, Ithaca, NY, 14853 USA, and with the  Institute of Electrical Engineering and Center for Neuroprosthetics, EPFL, 1202 Geneva, Switzerland (e-mail: us52@cornell.edu).}
\thanks{M. Shoaran is with the  Institute of Electrical Engineering and Center for Neuroprosthetics, EPFL, 1202 Geneva, Switzerland (e-mail: mahsa.shoaran@epfl.ch).}
}

\maketitle

\IEEEpeerreviewmaketitle

\begin{abstract}
The application of closed-loop approaches in systems neuroscience and therapeutic stimulation holds great promise for revolutionizing our understanding of the brain and for developing novel neuromodulation therapies to restore lost functions. Neural prostheses capable of multi-channel neural recording, on-site signal processing, rapid symptom detection, and closed-loop  stimulation are critical to enabling such novel treatments. However, the existing closed-loop neuromodulation devices are too simplistic and lack sufficient on-chip processing and intelligence. 
In this paper, we first discuss both commercial and investigational closed-loop  neuromodulation devices for brain disorders. Next, we review state-of-the-art  neural prostheses with on-chip machine learning, focusing on application-specific integrated circuits (ASIC). System requirements,  performance and hardware comparisons, design trade-offs, and hardware optimization techniques are discussed. \textcolor{black}{To facilitate a fair comparison and guide design choices among various on-chip classifiers, we propose a new energy-area (E-A) efficiency figure of merit that evaluates hardware efficiency and multi-channel scalability.}
Finally, we present several techniques to improve the key design metrics of tree-based on-chip classifiers, both in the context of ensemble methods and oblique structures.  A novel Depth-Variant Tree Ensemble (DVTE) is proposed to reduce  processing latency (e.g., by 2.5$\times$ on seizure detection task). We further develop a  cost-aware learning approach to jointly optimize the power and latency metrics.
We show that algorithm-hardware co-design enables the energy- and memory-optimized design of tree-based models, while preserving a high accuracy and low latency. 
Furthermore, we show  that our proposed tree-based models feature a highly interpretable decision process that is essential for safety-critical applications such as closed-loop stimulation.
\end{abstract}


%
\IEEEpeerreviewmaketitle

 \vspace{-2mm}
\section{Introduction}  \vspace{-1mm}
\IEEEPARstart{D}\noindent eveloping novel non-pharmacological treatments such as neurostimulation is becoming increasingly important to treat some of the most prevalent and intractable neurological disorders. Brain stimulation is currently the most common surgical treatment  for movement disorders and has shown promise in  epilepsy, neuropsychiatric disorders, memory, chronic pain, and traumatic brain injury, with new applications rapidly emerging. Despite promising  proof-of-concept results, current clinical neurostimulators are limited in many aspects.
For example, while deep-brain stimulation (DBS) can effectively control motor symptoms in most patients suffering from Parkinson's disease (PD), 
it causes persistent side effects (e.g., speech impairment and cognitive symptoms)  \cite{little2013adaptive, yao2020improved}. It is now widely known that this is due to the conventional ‘open-loop’ approach, which involves delivering constant high-frequency ($\sim$130Hz) stimulation regardless of the patient's clinical state. In addition, open-loop stimulation increases the power consumption  and the need for surgical battery replacement. This simplistic  open-loop approach is also a key limiting factor in designing clinically effective stimulation for more complex disorders such as depression \cite{lo2017closed}, Alzheimer's disease  \cite{ezzyat2018closed},  and stroke \cite{iturrate2018closed, huang2008cortical}, among others \cite{lozano2019deep, krauss2020technology, iturrate2018closed}. 

To further leverage the benefits of stimulation and address the aforementioned limitations, closed-loop neuromodulation techniques have been recently explored, such as the responsive neurostimulator for epilepsy \cite{skarpaas2009intracranial} and PD \cite{meidahl2017adaptive}, with promising results. In this approach, stimulation is dynamically controlled according to a patient's clinical state, either with a continuous (i.e., adaptive)  or an on-off (i.e., on-demand) strategy. Through feedback from relevant biomarkers of a neurological symptom (e.g., a seizure event, tremor episode, or mood change), closed-loop stimulation can titrate charge delivery to the brain, thus reducing the side effects and the amount of stimulation delivered, enhancing  the therapeutic efficacy and battery life compared to its open-loop counterparts \cite{yao2020improved}. However, several critical challenges remain to be addressed in order to fully exploit the potential of closed-loop therapies for neurological  disorders. The existing closed-loop devices mainly rely on simple comparison of a pre-selected biomarker (typically  from 1 out of 4 channels) against a fixed threshold. Such simplistic  approaches are known to be suboptimal in terms of predictive accuracy, resulting in low sensitivity and high false alarm rates, while exacerbating other symptoms \cite{krauss2020technology}. Multiple biomarkers and control loops may be necessary to reliably improve symptoms, leading to design complexity. 

\begin{figure}[t]
  \centering
  \includegraphics[width=0.9\columnwidth]{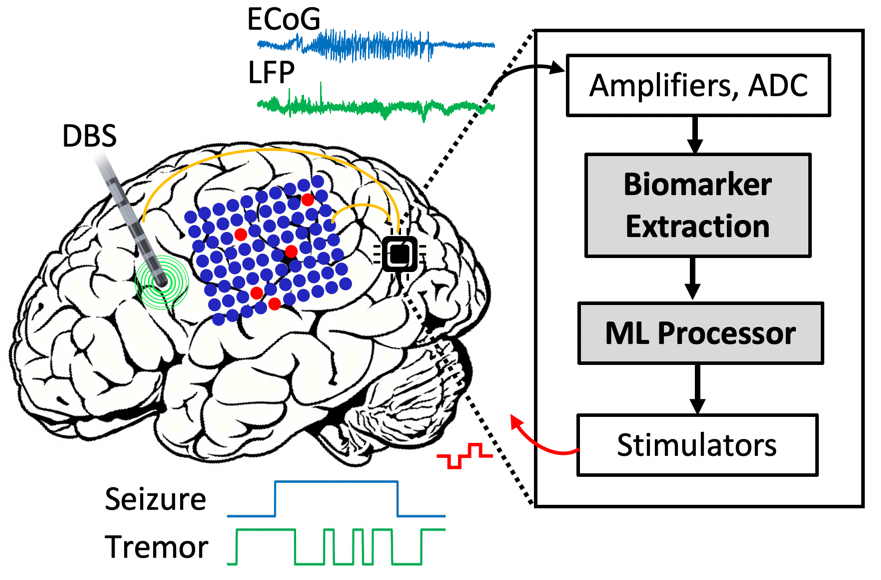}\vspace{-1mm}
  \caption{Symbolic view of a closed-loop neural prosthesis. Multi-channel neural signals such as ECoG and LFP are recorded by cortical and deep-brain electrodes and sent to the implantable microchip. The on-chip biomarker extraction and ML processor  detect the onset of  symptoms and trigger a therapeutic neurostimulator.}\vspace{-5mm}
  \label{block}
\end{figure}

A promising solution to address this challenge is to implement a machine learning (ML) algorithm directly on the implant or wearable to predict the onset or severity of neurological symptoms, an approach that has gained significant  interest in recent years \cite{yoo20128, shoaran2018energy, altaf201516, altaf20151,  kassiri2017closed, o2018nurip, chen2014fully, taufique2021low}. 
Real-time symptom control  can be achieved through on-chip biomarker extraction and ML-based disease state detection, followed by a closed-loop intervention (e.g., electrical, magnetic or optical stimulation, drug delivery) to suppress the abnormal activity, as illustrated in Fig. \ref{block}. This approach offers significant advantages over the conventional wireless transmission and external processing methods \cite{verma2010micro, zhang2010implantable} that suffer from feedback loop latency, high power consumption due to  continuous telemetry, security and privacy concerns \cite{zhan2019resource, zhu2020closed}. A number of clinical trials  have recently shown the advantage of machine learning-based control for closed-loop stimulation in movement disorders, epilepsy, and memory \cite{he2021closed, ezzyat2018closed}.  
\textcolor{black}{In addition, machine learning systems have been developed to forecast the onset of neurological symptoms  during preictal phase, allowing sufficient time  prior to seizure manifestation (e.g., in the order of several minutes) to provide early warnings to the patients and caregivers \cite{kuhlmann2018epilepsyecosystem, truong2019semi, dilorenzo2019neural}. In closed-loop neural prostheses, however, both the ML decoder and neurostimulator are integrated on the implant,  eliminating the need for excessively long symptom prediction horizons \cite{maiwald2004comparison}.
Therefore, most closed-loop devices train the classifier to differentiate ictal epochs from interictal period, several seconds prior to symptom onset \cite{jouny2011improving}. Such systems detect the onset and termination (i.e., offset) of neurological symptoms to precisely control the delivery of  stimulation  \cite{altaf201516}.}

Despite the benefits of using machine learning for closed-loop intervention, strict power and area requirements on an implantable or wearable device pose critical challenges for  hardware realization of ML algorithms, particularly in the form of a miniaturized ASIC. The choice of  learning algorithm and neural biomarkers affects the prediction accuracy and latency. Moreover, the  prediction accuracy  depends on the spatial resolution of the recording system and  the number of input channels.
Thus, there is a crucial need to develop high-performance, energy- and area-efficient biomarker extraction and ML solutions that are scalable to high channel counts and satisfy the implantable/wearable power budget and form factor. 

In this paper, we review the state-of-the-art neural prostheses with embedded biomarker extraction and machine learning. We first discuss the closed-loop system components, requirements for the next-generation smart neural prostheses, their clinical applications,  hardware techniques and trade-offs. Commercial and investigational closed-loop neuromodulation devices and a comparison of previously reported system-on-chips (SoCs) for neural signal classification are presented.
In the second part of this paper, we discuss an emerging class of machine learning algorithms based on decision trees \cite{shoaran2018energy, o202026, yang2017hardware, zhu2020resot, zhu2020closed}, including tree ensembles and oblique trees, that  are particularly suitable for energy- and area-constrained platforms such as brain implants and wearables. 
We introduce novel techniques to improve the accuracy-latency trade-off in tree ensembles.
A new class of tree-based models that effectively combine decision trees (DTs) with neural networks is further discussed. After presenting various techniques  for energy, latency,  and memory-efficient realization of oblique trees, 
we present the results of testing these models on two  neural signal classification tasks relevant to closed-loop stimulation (epilepsy and PD). 
 

It should be noted that closed-loop neural prostheses with on-chip intelligence are also being explored in the context of fully implantable  brain-machine interfaces (BMI) \cite{chen2015128, shaikh2019towards, yang2017hardware, do2018area, chae2009128}. Such BMI systems can provide a sensory feedback to the brain and/or control prosthetic devices to restore lost motor or sensory function in paralyzed patients.  However, the focus of this paper is on neural prostheses that directly record and modulate the brain activity to treat neurological disorders, while motor neuroprosthetics (i.e., BMIs or brain-computer interfaces, BCI), peripheral \cite{navarro2005critical} and spinal cord prostheses \cite{courtine2019spinal} (e.g., EMG-based interfaces) are beyond the scope of this paper. 
Furthermore, we limit our review to those systems that focus on ASIC implementation of neurological symptom detection algorithms (either validated on, or with a potential for closed-loop stimulation) due to similarity in design requirements. Thus,  FPGA-based systems are not included in this review. 
While the focus of this review is on CMOS-based edge machine learning specifically for neural prostheses, a comprehensive review on embedded hardware (FPGA, neuromorphic, CMOS) for neural networks used in biomedical applications can be found in \cite{rahimiazghadi2020hardware}.


This paper is an extension of our conference paper \cite{zhu2020closed} that presented a brief survey on closed-loop neural interface systems with on-chip machine learning  and provides the following contributions: 
\begin{itemize}
\item A comprehensive review on the latest developments in technology design for closed-loop stimulation, including novel electrodes for sensing and stimulation, emerging clinical applications,  commercial, research-based and investigational devices for closed-loop stimulation. 
\item A detailed review of the reported neural interface SoCs with on-chip machine learning for neurological disease detection, either as a stand-alone chip or as part of a closed-loop  system (implantable and wearable). 
\item \textcolor{black}{Future directions for the next-generation closed-loop neural prostheses, including the integration of advanced design techniques, accommodating high  channel counts and the need for online learning.}
\item Novel algorithm-hardware co-design techniques for  next-generation  energy-efficient neural prostheses. Specifically, we present  a range of methods for cost-aware implementation of  tree-based  classifiers in brain implants and validate them on human neurophysiological datasets. 
\end{itemize}

\begin{figure}[t]
	\centering
	\includegraphics[width=1\columnwidth]{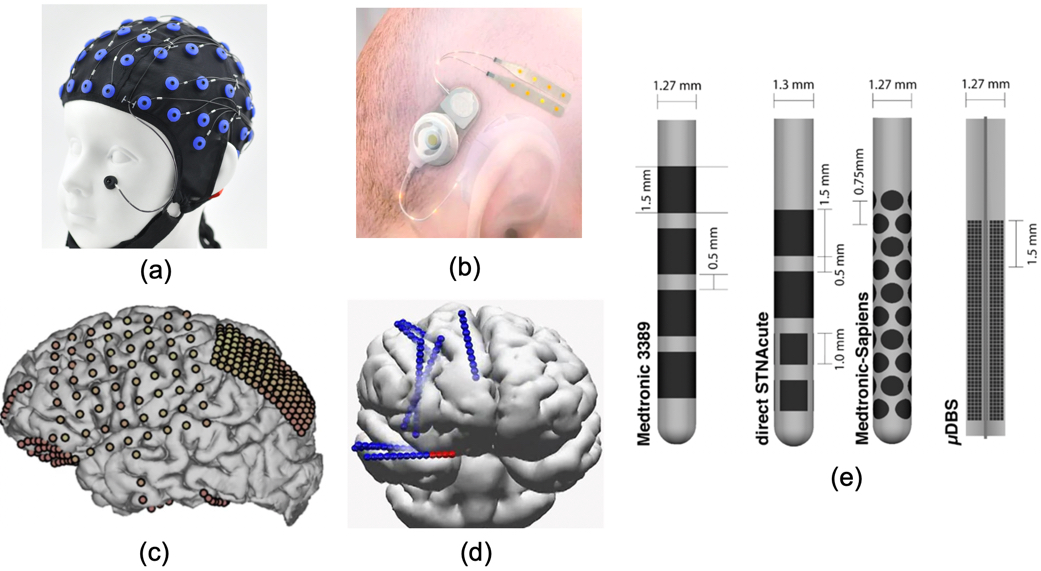}
	\caption{Standard and emerging electrodes for neural recording and stimulation via noninvasive, minimally-invasive, and invasive technologies; (a) Standard scalp-EEG electrodes. (b) The Epios subscalp EEG device for chronic epilepsy monitoring  \cite{duun2020new}. (c) Standard and high-density ECoG  \cite{parvizi2018promises}. (d) Stereo-EEG leads \cite{herff2020potential}. 
	(e) Clinical DBS (Medtronic's FDA-approved 3389, left), emerging directional DBS leads (8-channel direct STNAcute and 40-channel Medtronic-Sapiens, middle) and the Willsie and Dorval 1760-contact micro-DBS lead (right)  \cite{anderson2018optimized}. 
	} \vspace{-3mm}
	\label{fig-elec}
\end{figure}

\vspace{-2mm}
\section{Closed-loop Neural Prostheses: Recent Trends, System Requirements, and Trade-offs} 
In  a closed-loop neural prosthesis (Fig. \ref{block}), neurostimulation is triggered to suppress the impending signs of a neurological disease. 
Research on closed-loop neurostimulation has gained momentum in recent years, particularly with the success of proof-of-concept studies on epilepsy \cite{morrell2011responsive} and PD \cite{rosin2011closed, little2013adaptive, arlotti2016adaptive}. Closed-loop approaches are now being explored to treat a variety of medication refractory brain disorders where open-loop stimulation has been less effective. Yet, major technological challenges have limited the efficacy and clinical translation. These challenges include  the low channel count  of the current devices, the effect of stimulation artifacts on the sensing circuits, the need for miniaturization and improved energy efficiency, and the need for more advanced control algorithms \cite{yao2020improved, krauss2020technology, zhu2020closed}. Next-generation closed-loop neuromodulation systems will  require significant improvements in the existing devices. For instance,  higher numbers of recording and stimulation channels 
will be necessary for disorders that require multi-site neural recording and manipulation. More sophisticated processing algorithms and complex stimulation patterns will be beneficial to improve therapeutic outcomes. However, this will  increase  the design complexity and required on-chip resources for symptom detection and stimulation, as well as the required processing time. 
Better localization of target regions for effective stimulation and improved stimulation artifact cancellation are also critical for bidirectional neural prostheses.  In this paper, we discuss the major challenges and review the most recent advances in the field, with a particular focus on machine learning-embedded  implantable and wearable systems.   
\vspace{-2mm}
\subsection{Sensing and Stimulation}
High-density neural recording and multi-site neurostimulation 
with low-power miniaturized circuits are crucial for the next-generation closed-loop neural prostheses.
\textcolor{black}{Particularly, complex disorders such as depression and Alzheimer's disease (AD) 
need multi-site rather than single-site recording that calls for more intelligent, data-driven closed-loop systems with high-density sensing and stimulation capabilities.}

\subsubsection{Conventional and Emerging  Electrodes for Sensing and Stimulation}
In a neural prosthesis, the electrophysiological activity of the brain  can be recorded through various noninvasive, minimally-invasive, or invasive electrodes such as  scalp EEG, subscalp EEG \cite{duun2020new}, electrocorticography (ECoG), also known as intracranial EEG (iEEG), stereo-EEG (sEEG) \cite{mullin2016seeg, herff2020potential}, and deep-brain leads, providing various degrees of spatial and temporal resolution (Fig. \ref{fig-elec}). In some cases and predominantly in implantable prostheses, the same electrode can be used for delivering electrical stimulation to the brain to suppress disease symptoms. 

The EEG electrodes have a cm-range distance and are noninvasive. Both scalp  and subscalp EEG are suitable for wearable settings, with electrodes placed either above (scalp EEG) or under the scalp (subscalp EEG).
Subscalp electrodes are particularly suitable for chronic (i.e., longer than one month) EEG recording in a home environment and require a minimally invasive surgery under general anesthesia to implant the subcutaneous electrodes    \cite{duun2020new}.  The subscalp approach eliminates the need for constant electrode care (i.e., no need for an EEG cap or adhesives electrodes), providing a stable and less  obtrusive recording modality compared to conventional EEG, Fig. \ref{fig-elec}(b). Furthermore, subscalp EEG has been shown to attenuate several types of artifacts and improve (or at least maintain) the signal quality compared to EEG. However, similar to scalp EEG, it is limited in temporal and spatial resolution compared to  ECoG (i.e., $<$100Hz vs. several hundred Hz) and  cannot monitor deep-brain structures. A number of subscalp EEG  systems  are currently certified or  in development for long-term epilepsy monitoring (Section III). 

The spacing of ECoG electrodes (epidural or subdural) is typically within mm-range, while  state-of-the-art  ECoG interfaces  enable denser recording arrays for high-spatial-resolution recording of cortical activity \cite{viventi2011flexible}. For instance, it has been shown that high-density $\mu$ECoG with a 400$\mu$m pitch outperforms lower density grids in classifying cognitive tasks in humans \cite{hermiz2018sub}, highlighting its potential for future high-performance neuroprosthetic applications. High-frequency electrophysiological activity relevant to seizure prediction or epileptic foci localization can be captured on high-resolution ECoG  from submillimeter scale cortical regions \cite{stead2010microseizures, viventi2011flexible, shoaran2014compact, shoaran2012design}.
These novel  electrodes are not yet adopted in  diagnostic or closed-loop devices.

While ECoG provides a precise mapping technique at the level of cortical surface, stereo-EEG (sEEG) \cite{herff2020potential} is an alternative  minimally-invasive method for identifying seizure onset zone in medically refractory focal epilepsy. Placement of stereo-EEG electrodes (typically 5--15 cylindrical shafts) requires small, localized burr holes to insert depth electrodes into the brain.  Stereo-EEG enables a sparse sampling of localized  brain regions, as opposed to the relatively large craniotomy required for strip/grid ECoG implantation \cite{herff2020potential}.

The electrodes on a deep-brain lead (e.g., Medtronic 3387/3389 deep-brain stimulation lead  with four cylindrical contacts) are placed several millimeters or even 100s of micrometers apart to capture the local field potential (LFP) activity (up to several 100 Hz) \cite{anderson2018optimized}. 
The leads employed in sEEG are similar to those used for deep brain stimulation (DBS). DBS is widely used as a treatment for essential tremor, PD and dystonia, with emerging applications in epilepsy, major depression, obsessive-compulsive disorder (OCD), and Tourette's syndrome. While DBS is primarily used for electrical stimulation, the  chronic efficacy and stability of DBS leads suggest the use of long-term sEEG for sensing applications and closed-loop prostheses \cite{herff2020potential}. In rare cases, single-unit activity captured by $\mu$DBS leads (100$\mu$m spacing \cite{anderson2018optimized})  or penetrating microelectrodes such as  Utah array can be used to detect spike-based biomarkers (e.g., neuronal firing rates correlating with cognitive functions) for disease state prediction and guiding neurostimulation therapy \cite{baker2016robust, guggenmos2013restoration}.

For stimulation, recent DBS electrodes employ directional leads with higher number of small contacts (e.g., 16, 40, 1760) as opposed to traditional leads with  only four cylindrical contacts \cite{anderson2018optimized, krauss2020technology}, Fig. \ref{fig-elec}(e). Such directional leads with segmented electrodes can effectively steer the stimulation back toward a missed target structure,  without exciting non-target regions and inducing adverse effects. 
Moreover, recent studies report the impact of using temporal patterns delivered via multiple contacts in enhancing plasticity and symptom relief \cite{anderson2018optimized, krauss2020technology}, highlighting the benefits of high-channel-count stimulation. 


\subsubsection{Concurrent Sensing and Stimulation}
Accuracy and latency can be enhanced by measuring evolving disease state even as therapeutic stimulation is  applied. This motivates the need for a new class of circuit and system techniques to enable detection of weak electrophysiological signals of interest in the presence of orders-of-magnitude stronger stimulus artifacts. This general problem of measuring weak signals in the presence of extreme self-interference represents a general challenge for modern mixed-signal circuit in various sensing and communication applications. 
The next generation ‘full-duplex’ neuromodulation devices must feature simultaneous sensing and stimulation for truly closed-loop operation. 

The most common electrical approach is to use ‘blanking’ \cite{demichele2003stimulus, blum2007integrated} where recording amplifiers are disconnected from the electrode during and immediately after stimulation, and then reconnected after the stimulation artifact will no longer saturate the amplifier. Recent improvements allow the amplifier to be connected immediately
after stimulation \cite{johnson2017implantable}, using mixed-signal circuit realization of the analog front-end (AFE). However, this method still suffers from its inability to record while stimulating, which is especially limiting in complex, multi-electrode stimulation patterns where extended stimulation blocks recording over much longer time stretches. 

An alternative approach based on high dynamic range (DR) AFE incorporating amplifiers and analog-to-digital converters (ADC) can reliably record the neural signal along with the persistent artifacts without saturation \cite{rozgic20180, chandrakumar201780}. 
Alternatively, the design in \cite{zhou2019wireless} proposes a linear-interpolation-based artifact cancellation implemented on an FPGA. 
%
Another approach employs a front-end cancellation technique that avoids using a high DR AFE \cite{mendrela2016bidirectional}. However, this method requires a significant convergence time (impractical for closed-loop systems). 
Artifact cancellation generally poses additional hardware overhead on the AFE and on the back-end for digital cancellation, which limits the area and energy efficiency of the  closed-loop system. 
\vspace{-4mm}
\subsection{Disease Biomarkers and Machine Learning} \vspace{-1mm}
While artificial intelligence and machine learning can contribute to various aspects of neurotechnology (e.g., optimizing the programming of stimulation to activate target regions, offline analysis of chronic neural recordings, understanding the underlying disease mechanism), our focus in this paper is on real-time on-device disease state prediction using machine learning. This is inspired by the unique potential of ML techniques in classifying high-dimensional electrophysiological signals, typically outperforming conventional methods in various applications \cite{shoeb2010application, yoo20128, shoaran2018energy, yao2020improved, zhang2014seizure, zhu2019migraine, yao2021predicting}.
Accurate and timely detection of symptoms in brain disorders is critical to enable closed-loop neuromodulation, and it typically requires  the use of correlating biomarkers (i.e., features) of an underlying disease state along with a machine learning algorithm.
The widely used features in electrophysiological studies  include the spectral power (or bandpower) in various frequency bands relevant to the neurological symptom of interest, time-domain and statistical features (e.g., line-length \cite{esteller2001line}, the Hjorth parameters of activity, mobility, and complexity \cite{hjorth1970eeg, yao2020improved, zhu2019migraine}, number of peaks, peak-to-peak amplitude and peak latency \cite{zhu2019migraine}), biomarkers that measure  connectivity between different brain regions  such as phase-amplitude coupling and phase locking value \cite{abdelhalim201364, yao2018resting, yao2020improved, yao2021predicting, yao2020mental}, 
and the correlation structure of multi-channel neural data \cite{schindler2007assessing}.

Some initial steps have been taken recently toward embedding biomarkers and machine learning  algorithms on brain implants  or wearables for disease monitoring and closed-loop therapy,  and in investigational neuromodulation systems such as Medtronic's Summit RC+S \cite{stanslaski2018chronically} and Percept PC systems, as summarized in the next \textcolor{black}{sections}. 


\subsubsection{Classifier requirements -- High accuracy, low latency}  
Symptom detection requires high accuracy and low latency. The classification algorithms should be robust in handling the typically small amounts of training data in such applications, due to the lack of chronic recordings. In some cases, the recording length could be limited to the duration of surgery for device implantation (e.g., up to 30 minutes for DBS surgery in PD, several days for epilepsy patients undergoing pre-surgery evaluation at the hospital). With the increasing interest in devices with chronic recording capability (e.g., the NeuroPace RNS and Medtronic Percept), it is expected that more long-term human data will be available in near future, enabling data-driven  algorithm and hardware developments.

Depending on the  distribution of different classes in a neurophysiological dataset, the appropriate measure of accuracy may be used to evaluate the classifier's performance. Sensitivity (i.e., True Positive rate), specificity (i.e., selectivity or True Negative rate), accuracy,  F1 score,  the area under the ROC curve (AUC), and the false alarm rate (FAR) are among the commonly used metrics in ML studies on neural datasets. 
The F1 score (i.e., the harmonic mean of sensitivity and precision: 2$\times$(precision$\times$sensitivity)/(precision+sensitivity), where precision represents the positive predictive value)  
is particularly  useful  in dealing with imbalanced datasets (i.e., datasets with non-uniform distribution of classes), such as EEG or  iEEG recordings in epilepsy \cite{shoaran2018energy}. Balanced accuracy  (i.e., the average of sensitivity and specificity) is another metric used for  imbalanced datasets \cite{yao2021predicting}.

Most closed-loop systems rely on external computing for feature extraction and classification, which suffers from long loop latency, thus jeopardizing the real-time feedback.  The on-chip integration of ML can significantly speed up the closed-loop therapy and enable feedback loops of msec-range latency.
If the feedback is  too slow, the  detector may miss the window of opportunity to trigger or adjust stimulation, resulting in poor therapeutic outcomes.
More sophisticated processing algorithms  may improve the decoding accuracy at the cost of increased processing latency. 

While `latency' has been used to represent various types of `processing delay' in literature (e.g., feature extraction and classification delay resulting from window-based processing),  the detection latency of a closed-loop system is typically defined as the delay between  the electrographic, expert-marked, or externally labeled symptom onset and the onset declared by the on-chip processor, for instance in detecting seizures in epilepsy \cite{shoeb2010application, shoaran201616, shoaran2018energy, altaf20131, zhu2021unsupervised, page2014flexible, shoaran2015fully} or tremor onset in PD \cite{yao2018resting, yao2020improved, wang2018towards}. In disorders such as epilepsy, the  onset of clinical symptoms could be several  seconds (in some cases, up to 30 seconds
\cite{jouny2011improving}) after the time of earliest detectable changes in  neural activity.  Therefore, therapeutic feedbacks within that time frame can be still beneficial for the patients. In other cases, e.g., in movement disorders with more rapid changes in electrophysiological state, a low latency  (i.e., negative latency or  lead \cite{yao2020improved}) is preferred to enable closed-loop stimulation.



\subsubsection{Classifier requirements -- Low power and small area}
To enable efficient local processing in a brain implant,  silicon-realizable ML algorithms that can precisely predict a neurological symptom 
are essential. 
Neural prostheses with on-device ML do not require continuous wireless telemetry. Yet, low-power realization of machine learning algorithms  is crucial to avoid excessive power dissipation. Optimized use of memory and computational resources and compact silicon area are further required to process multiple channels.
The computational complexity of the classifier (and features) could set a limit on the number of input channels, thus hindering its application  in more complex disorders.

The conventional implementation of most classification algorithms is resource intensive  such that devices in existence today \cite{morrell2011responsive} sacrifice the  classification accuracy and latency to meet the power and size constraints \cite{shoaran2018energy}. Some limited processing is embedded in recently developed neuromodulation devices, but this applies to 1--4 channels only, requiring external classifiers for more accurate symptom detection \cite{stanslaski2018chronically}. There is a crucial need for energy- and area-efficient machine learning algorithms via co-design of algorithm and hardware, as discussed in the next \textcolor{black}{sections}.


\begin{figure}[t]
	\centering
	\includegraphics[width=1\columnwidth]{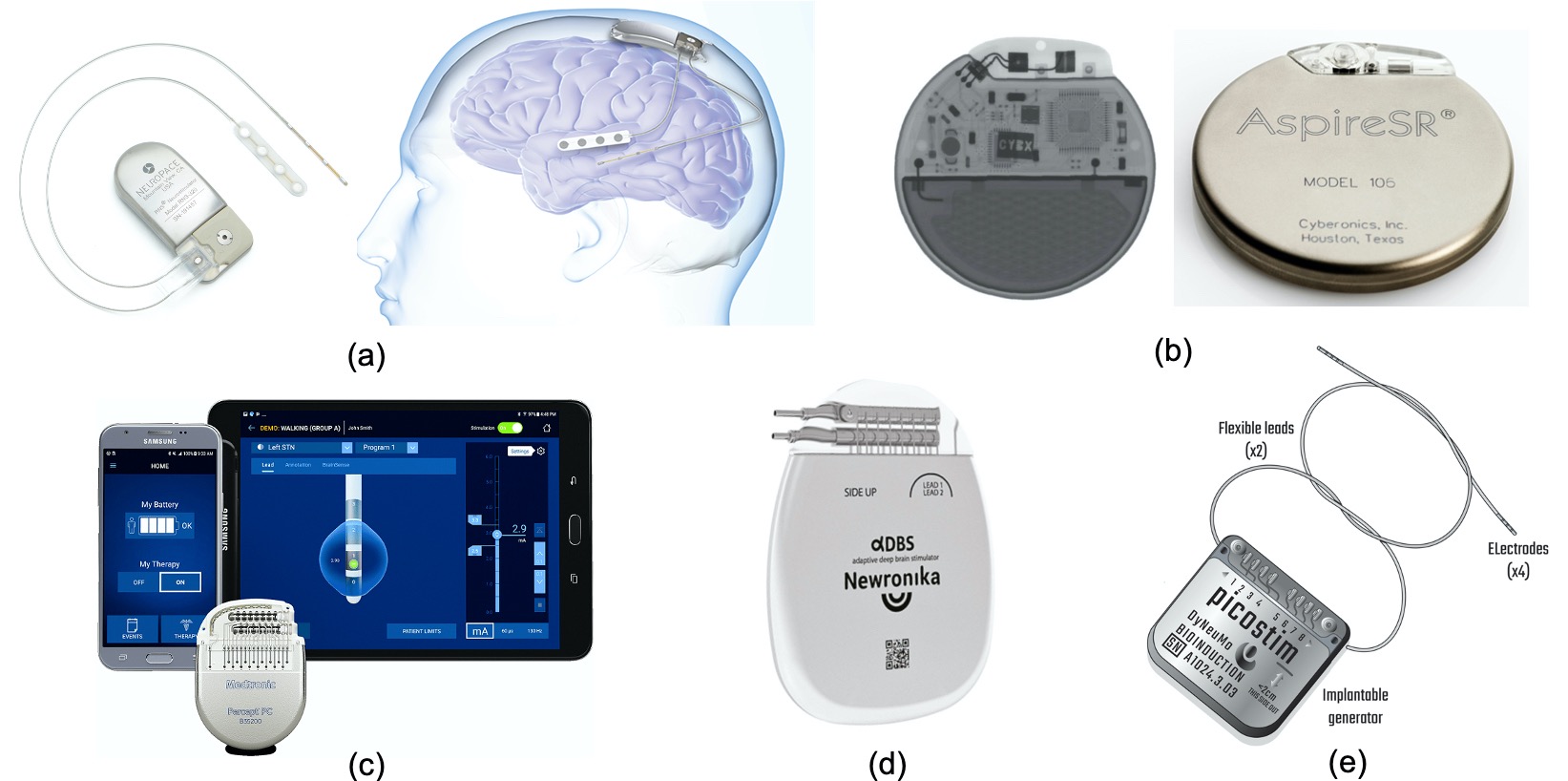}
	\caption{Existing clinical or research-based closed-loop neuromodulation devices (with or without on-device ML);
		(a) The NeuroPace RNS device for epilepsy. (b) The AspireSR (Cyberonics, now known as LivaNova) device for epilepsy. (c) The Medtronic Percept PC device for movement disorders. (d) The Newronika AlphaDBS system for Parkinson's disease. (e) The DyNeuMo Mk-1 system for movement disorders. 
	}\vspace{-3mm}
	\label{fig-dev}
\end{figure}

\subsubsection{Neurophysiological Datasets}  
In contrast to computer vision tasks that benefit from standard datasets for direct benchmarking of  machine learning models,
the electrophysiological datasets used in disease prediction tasks are diverse and not directly comparable. Furthermore,  these datasets include different numbers of patients with various levels of symptom detection complexity, making it challenging to compare  classifiers evaluated on the same dataset but on different patients.
Another critical challenge is the lack of data sharing and open-source datasets in emerging applications beyond epilepsy (e.g., movement disorders, depression,  Alzheimer's disease), which greatly limits the development of biomarkers and ML solutions and subsequent device implementation for such disorders.
\vspace{-1mm}
\section{Commercial and Investigational \\Closed-loop Devices} \vspace{-1mm}

One of the few  platforms currently available for closed-loop stimulation is the NeuroPace's Responsive Neurostimulator (RNS) for medication-refractory epilepsy  (Fig. \ref{fig-dev}(a)). RNS continuously analyzes cortical activity to detect and halt seizure events from 4 channels, by comparing a simple pre-selected feature (signal intensity, line-length, or half-wave) against a threshold \cite{morrell2011responsive, bergey2015long}, and it is currently in clinical use in  patients. Both cortical and deep-brain stimulation are enabled in RNS (8  channels).
The recently published results of a nine-year, multi-center chronic  study of RNS device  on 230 patients in 34 epilepsy centers \cite{nair2020nine} showed significant  reductions in seizure rates: 75\% median reduction,   at least 50\% reduction in 73\% of patients. 
The sudden unexpected death in epilepsy (SUDEP)  was also significantly reduced. The responsive neurostimulation was a well-tolerated  treatment, with a similar safety profile to other epilepsy procedures. 


Similarly, Medtronic's investigational Activa PC+S, Summit  RC+S \cite{stanslaski2018chronically} and Percept PC system (Fig. \ref{fig-dev}(c)) are capable of sensing and closed-loop stimulation for movement disorders such as essential tremor and PD. Compared to RNS, the Medtronic devices implement slightly more complex spectral analysis and a linear classifier, relying on only 4 sensing channels with 2--8 features in total, and 8--16 stimulation channels. 
For both RNS and Medtronic devices, external algorithms with  advanced machine learning capabilities may be necessary for more accurate symptom tracking \cite{stanslaski2018chronically, jarosiewicz2020rns}, at the cost of long loop latency and high power demands to support continuous wireless streaming \cite{shoaran2018energy, zhu2020resot}. 
The AspireSR 106 (LivaNova) is an implantable Vagus Nerve Stimulator (VNS) with an optional  AutoStim mode in which  the VN stimulation can be adjusted in response to ictal heart rate changes which are potentially associated with an impending seizure (Fig. \ref{fig-dev}(b))  \cite{hamilton2018clinical}. 
%
In a  study on the efficacy of open-loop VNS on 5554 patients  \cite{englot2016rates}, a growing increase  in seizure freedom was observed  post therapy, with  49\%  responding to treatment  0--4 months after implantation (i.e., $>$50\% seizure frequency reduction). The efficacy of closed-loop AspireSR versus the preceding open-loop device was recently studied, where 4 (from 11) patients  who were less responsive to the open-loop VNS achieved $>$50\% seizure reduction  \cite{kawaji2020additional}. Of note, there have been reports on the Federal Drug Administration (FDA) device recall for different models of VNS due to concerns on device malfunctions.

In addition to the devices described above, there is an increasing effort in developing novel closed-loop stimulation devices for a variety of brain disorders. One example is the AlphaDBSTM system \cite{newronika}, which recently received the CE mark approval to treat Parkinson’s disease (Fig. \ref{fig-dev}(d)). This closed-loop system developed by Newronika  (S.p.A, Milan, Italy) can record deep-brain local field potentials and adjust the stimulation amplitude and frequency.
DyNeuMo (Bioinduction, Bristol, UK) is a  closed-loop neuromodulation research device that can  titrate stimulation according to the current motor state (e.g., posture and activity)  \cite{zamora2020dyneumo} (Fig. \ref{fig-dev}(e)). The device uses  off-the-shelf consumer technology and embeds three-axis accelerometer sensors and  8-channel programmable neurostimulators, and is currently in preparation for first-in-human research trials.


Minimally-invasive signal modalities such as subscalp EEG are also being considered for long-term epilepsy monitoring. For instance, the Epios device (Wyss Center for Bio and Neuroengineering, Geneva, Switzerland) \cite{duun2020new} enables both focal recording and full-montage coverage using subscalp EEG for chronic seizure analysis and forecasting (Fig. \ref{fig-elec}(b)). The EEG data is wirelessly transmitted to a  wearable unit and temporarily stored,  supporting multimodal ECG, audio, and accelerometry recording. Signals are   then transmitted to the cloud for long-term data analysis and visualization. The Epios device is currently in preparation for clinical trial phase. The  Minder device  (Epi-Minder, Melbourne, Australia) \cite{duun2020new} implants an electrode lead across the skull  to cover  both hemispheres (Fig. \ref{fig-dev}(g)). This subscalp system provides continuous long-term measurement of EEG for chronic epilepsy diagnosis  and monitoring (clinical trial  in progress). 
Alternatively, in the EASEE system  by Precisis (Heidelberg, Germany)  five subscalp  electrodes are implanted above  the seizure focus  for sensing and closed-loop neurostimulation with a personalized setting (clinical trial in progress) \cite{duun2020new}.

\vspace{-2mm}
\section{Neural Prostheses with On-chip ML}
In recent years, the application of machine learning  techniques  in closed-loop neuromodulation and  its CMOS implementation have received considerable interest.
Machine learning has been used to more accurately predict optimal stimulation times \cite{altaf201516, yao2020improved, o2018nurip, yao2021predicting, provenza2019decoding} and several clinical studies  have  shown the advantage of ML-based  closed-loop therapy in movement disorders \cite{he2021closed}, epilepsy \cite{kremen2018integrating}, and  memory \cite{ezzyat2018closed}.
The most prominent benefits of integrating machine learning algorithms on a brain implant include:
\begin{itemize}
	\item Eliminating the need for excessive wireless transmission for external processing, thus allowing design miniaturization, lower power dissipation, and higher mobility.

\item  Increasing patient independence and alleviating security concerns by avoiding the transmission of private data.

\item Improving  symptom prediction accuracy and latency.
\end{itemize}
The latter advantage largely depends on the number of sensing channels, the quality of neural signal (e.g., its sampling rate and signal-to-noise ratio), the choice of  machine learning algorithm and neural biomarkers, and the chronic robustness of the algorithm. As mentioned in the previous \textcolor{black}{section}, current clinical devices do not offer sufficient embedded biomarker extraction and ML, relying on telemetry and cloud-based processing for accurate symptom prediction.

Various hardware implementations of machine learning algorithms have been reported for  neurological symptom detection, as discussed below. 
Here, we limit our review to state-of-the-art neural prostheses with an ASIC implementation, validated  on animal or human datasets (acute and/or chronic, either diagnostic only or closed-loop).

\begin{figure*}[t] 
	\centering
	\includegraphics[width=2\columnwidth]{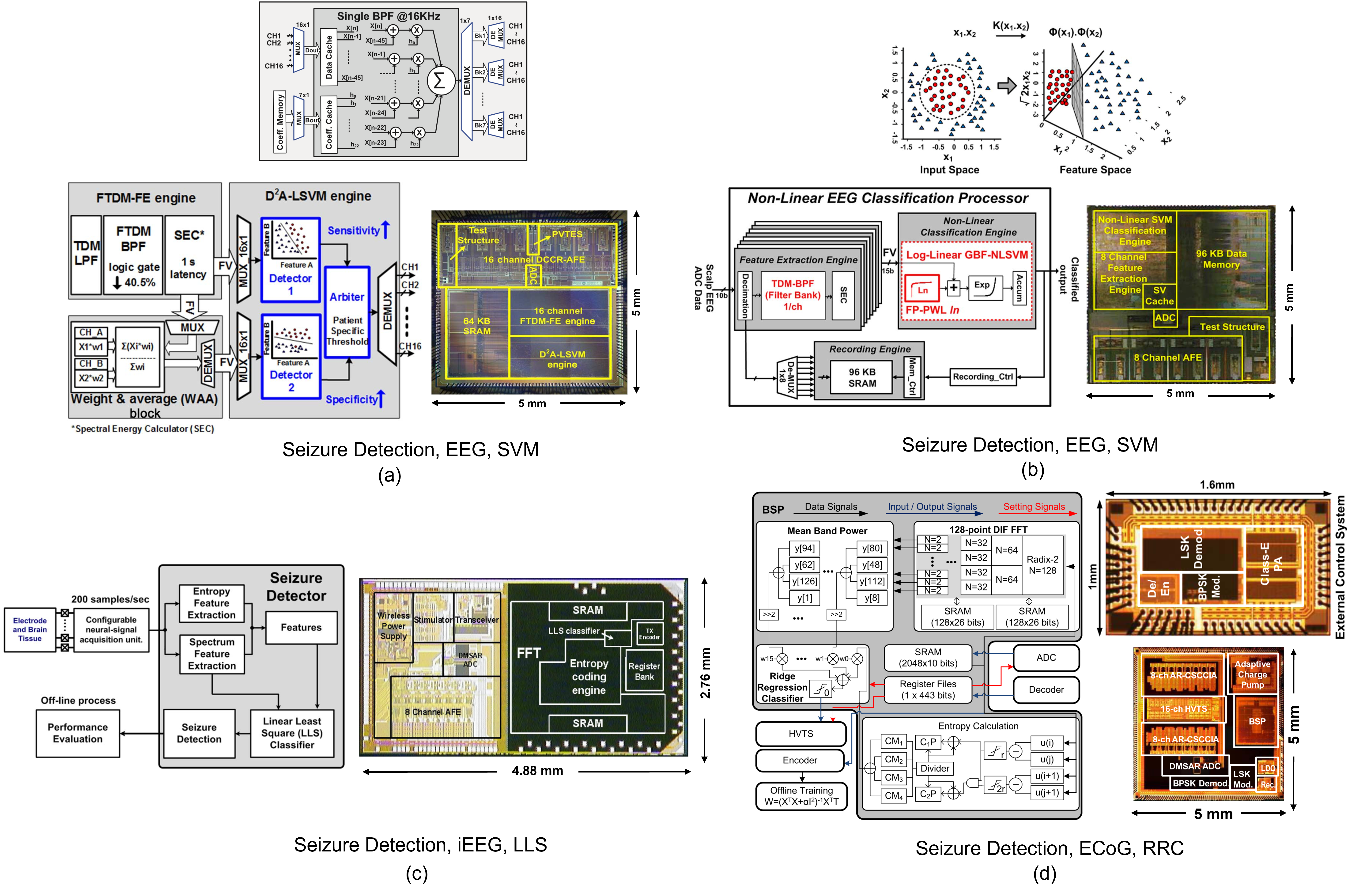}
	\vspace{-3mm}
	\caption{Hardware architectures and chip micrographs of ML-embedded neural prostheses for epilepsy: (a) Linear dual-detector  SVM classifier and closed-loop transcranial neurostimulator \cite{altaf201516}, (b)
		non-linear SVM-based  seizure detector \cite{altaf20151}, (c) linear least square (LLS) classifier and closed-loop stimulator \cite{chen2014fully}, (d) ridge regression classifier (RRC) and closed-loop stimulator~\cite{cheng2018fully}.}	
	\label{figsoc1}
\end{figure*}
\vspace{-2mm}

\begin{figure*}[t]\ContinuedFloat
	\centering
	\includegraphics[width=2\columnwidth]{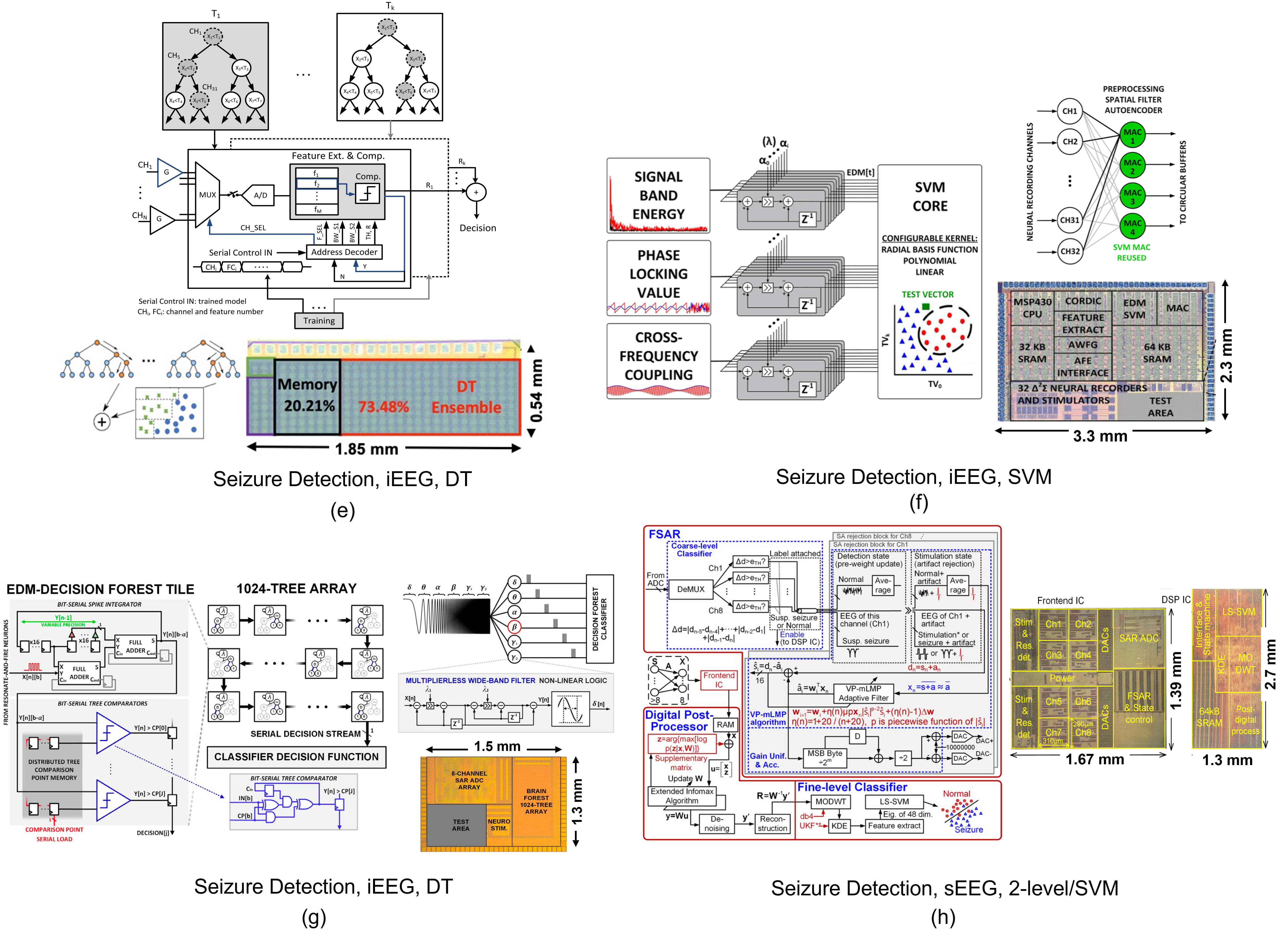}
	\vspace{-3mm}
	\captionsetup{labelfont={color=black},list=off,format=cont}
	\caption{Hardware architectures and chip micrographs of ML-embedded neural prostheses for epilepsy: (e) Gradient-boosted tree ensemble for seizure detection \cite{shoaran2018energy}, (f) exponentially decaying-memory SVM and closed-loop stimulator \cite{o2018nurip}, (g) AdaBoost decision tree classifier and closed-loop stimulator \cite{o202026}, (h) two-level coarse/fine classifier and closed-loop stimulator~\cite{wang202026}.}	
	\label{figsoc1}\vspace{-3mm}
\end{figure*}

\begin{table*}[t]
	\caption{Performance Summary of Machine Learning SoCs for Epilepsy}
	\vspace{-3mm}
	\begin{center}
\scalebox{0.64}{
		\begin{tabular}{l|c|c|c|c|c|c|c|c|c|c}
			\hline \hline
			Parameter & JETCAS'18 \cite{shoaran2018energy}  & JSSC'13 \cite{chen2014fully} & ISSCC'20 \cite{o202026} &  JSSC'18 \cite{cheng2018fully}  &  ISSCC'18 \cite{o2018nurip}  &  ISSCC'20 \cite{wang202026} & TBCAS'16 \cite{altaf20151} &  JSSC'13 \cite{yoo20128}  & JSSC'15 \cite{altaf201516} & \textcolor{black}{This Work} \\
			\hline \hline
			Process & 65 nm & 180 nm & 65 nm  & 180 nm & 130 nm & 180 nm & 180 nm & 180 nm & 180 nm & \textcolor{black}{65 nm} \\
			\hline
			Classifier & XGB DT  & LLS & AdaBoost DT  &   RRC  &   EDM-SVM  &  coarse/fine SVM & Non-Lin SVM   &   Lin-SVM  & Dual-LSVM & \textcolor{black}{DVTE$^+$} \\
			\hline
			Features & LLN, Pow, Var, BPF   & Ent., Spec.  &  RAF-BPF  &  FFT, ApEn    &   PLV, CFC, BPF  & MODWT-KDE  &   TDM-BPF &  BPF   & FTDM-BPF & \textcolor{black}{LLN, Var, BPF} \\
			\hline
			Signal Modality & iEEG  & iEEG & iEEG &  ECoG  &  iEEG  & Stereo-EEG & EEG  &  EEG   &   EEG & \textcolor{black}{iEEG} \\
			\hline
			Closed-loop & N  & Y & Y  & Y  &  Y &  Y &  N & N &  Y & \textcolor{black}{N} \\
			\hline
			\# of Sensing Channels & 32  & 8 & 8 & 16  & 32$^\dag$ & 8  & 8 & 8 & 16 & \textcolor{black}{32} \\
			\hline
			ML Energy Efficiency & 41.2 nJ/class. & 77.9 $\mu$J/class. & 36 nJ/class. & 62.5 $\mu$J/class. & 168.6 $\mu$J/class. & 14.2 $\mu$J/class. & 1.31 $\mu$J/class.$^{\dag\dag}$  &  1.49 $\mu$J/class.$^{\dag\dag}$ &  1.85 $\mu$J/class. & \textcolor{black}{\textbf{5.6 nJ/class.}} \\
			\hline
			ML Power & 206.4 $\mu$W & 882 $\mu$W$^{\ddag}$ & 9.6 $\mu$W$^{\dag\ast}$ & 2.5 mW$^{\ddag}$ & 674.4 $\mu$W & 1.16 $\mu$W & 156.6 $\mu$W$^{\ddag}$  &  193.8 $\mu$W$^{\ddag}$ &  216.7 $\mu$W$^{\ddag}$ & \textcolor{black}{\textbf{2.8 $\mu$W}} \\			
			\hline
			Total Area (ML Area) & 1 (1) mm$^2$ & 13.47 (4.85$^{\ast}$) mm$^2$ & 1.95  (0.42) mm$^2$ & 25  (2.52$^{\ast}$) mm$^2$ & 7.59  (3.32) mm$^2$  & 5.83  (3.51) mm$^2$ & 25  (5.55$^{\ast}$) mm$^2$ & 25  (7.37$^{\ast}$) mm$^2$ & 25  (7.47$^{\ast}$) mm$^2$ & \textcolor{black}{\textbf{0.31  (0.31) mm$^2$}} \\
			\hline
			Sampling Rate/Ch. & 5 kS/s & 62.5 kS/s & 256 S/s & 2 kS/s & 256 S/s & 1 kS/s$^{\ast\ast}$ & 128 S/s$^{\dag +}$ & 128 S/s$^{\dag +}$ & 128 S/s$^{\dag +}$ & \textcolor{black}{500 S/s} \\
			\hline
			Sensitivity & 83.7\% & 92\%$^{\P}$ & 96.7\% & 97.8\%$^{\P}$ & 97.7\% &  97.8\% & 95.1\% & 82.7\%$^{\P\P}$ & 95.7\% & \textcolor{black}{91.1\%} \\
			\hline
			Specificity & 88.1\% & N.A. & 0.8 FAR$^{\ast +}$ & N.A. &  0.185 FAR$^{\ast +}$ & 99.7\% & 0.27 FAR$^\S$  & 4.5\% FPR & 98\% (0.27 FAR$^{\ast +}$) & \textcolor{black}{96\%} \\
			\hline
			Dataset (\# patients) & iEEG.org (26) & Rats & EU-iEEG & ECoG (5) & EU-iEEG (4) & CHB-MIT$^{\S}$ (23) &  CHB-MIT (24) & CHB-MIT (24) & CHB-MIT (23) & \textcolor{black}{iEEG.org (11)} \\
			\hline			
			Latency & 1.1 s  & 0.8 s & N.A. & 0.76 s  & $<$0.1 s$^{\S\S}$  &  $<$0.3 s$^{\S\S}$  & 2 s & $<$2 s$^{++}$ & 1 s$^{\S\S}$ & \textcolor{black}{0.52 s$^{\S\S}$} \\
			\hline			
			ML Energy/Ch. & 1.29 nJ/S & 1.76 nJ/S & 4.69 nJ/S & 78.1 nJ/S & 82.3 nJ/S & 0.145 nJ/S & 153 nJ/S & 189 nJ/S & 106 nJ/S & \textcolor{black}{\textbf{0.175 nJ/S}} \\
			\hline			
			ML Area/Ch. & 0.031 mm$^2$ & 0.606 mm$^2$ & 0.053 mm$^2$ & 0.157 mm$^2$ & 0.104 mm$^2$ & 0.439 mm$^2$ & 0.694 mm$^2$ & 0.921 mm$^2$ & 0.467 mm$^2$ & \textcolor{black}{\textbf{0.01 mm$^2$}} \\
			\hline
			ML E-A FoM & 40.3 pJ$\cdot$mm$^2$/S & 1.07 nJ$\cdot$mm$^2$/S & 248 pJ$\cdot$mm$^2$/S & 12.3 nJ$\cdot$mm$^2$/S & 8.5 nJ$\cdot$mm$^2$/S & 63.6 pJ$\cdot$mm$^2$/S & 106.1 nJ$\cdot$mm$^2$/S & 174.3 nJ$\cdot$mm$^2$/S & 49.4 nJ$\cdot$mm$^2$/S & \textcolor{black}{\textbf{1.7 pJ$\cdot$mm$^2$/S}} \\
			\hline \hline
		\multicolumn{5}{l} {$\ddag$ ML (feature extractor and classifier) power consumption estimated from power breakdown} &			
		\multicolumn{5}{l}	{$\dag$ 4-channel post dimensionality reduction} \\
		\multicolumn{5}{l}	{$\dag\ast$ ML dynamic power (static power not reported)} &		
		\multicolumn{5}{l}	{$\dag\dag$ As reported in \cite{altaf201516}} \\
		\multicolumn{5}{l}  {$^+$ Also applicable to Parkinson tremor detection. Post place-and-route results.} &
		\multicolumn{5}{l}  {$\ast$ ML (feature extractor and classifier) area estimated from chip micrograph} \\
		\multicolumn{5}{l}	{$\ast\ast$ Variable (256, 1k, 125kS/s)} &
		\multicolumn{5}{l}	{$\P$ Accuracy metric} \\
		\multicolumn{5}{l}	{$\P\P$ Seizure detection rate} &
		\multicolumn{5}{l}	{$\ast +$ Number of false alarms per hour} \\
		\multicolumn{5}{l}	{$\S$ With 2000 seizure samples generated by synthetic minority oversampling technique} &
		\multicolumn{5}{l}	{$\S\S$ Processing (system) latency} \\
		\multicolumn{5}{l}	{$++$ Rapid eye blink detection} &
		\multicolumn{5}{l}	{$\dag +$ After on-chip decimation}\\
		\end{tabular}}
		\label{comparison1} 
	\end{center} 
\end{table*}

\vspace{-1mm}
\subsection{Implants and Wearables for Epilepsy} \vspace{-1mm}
The most common application of on-chip classification in a neural prosthesis is in the context of seizure detection for medically refractory epilepsy, where a supervised ML algorithm is typically used to detect the onset of seizure events from multi-channel neural recordings. 
Neurostimulation  offers an attractive treatment for intractable epilepsy (approximately one third of epileptic patients). Due to severity of refractory epilepsy, open-source epileptic EEG datasets (both scalp and intracranial) are largely available, as well as established animal models for device validation and preclinical studies. Therefore, several groups have integrated various biomarkers and machine learning algorithms on ASIC for automated seizure detection \cite{yoo20128,altaf20131,altaf20151,shoaran2018energy,lee2013low, huang20191, shoaran2017efficient, yang2020seizure , shoaib20140, taghavi201841} and  for controlling an on-chip stimulator \cite{chen2014fully, o2018nurip, wang202026, altaf201516, cheng2018fully, o202026}.

\begin{figure*}[t!]
	\centering
	\includegraphics[width=2\columnwidth]{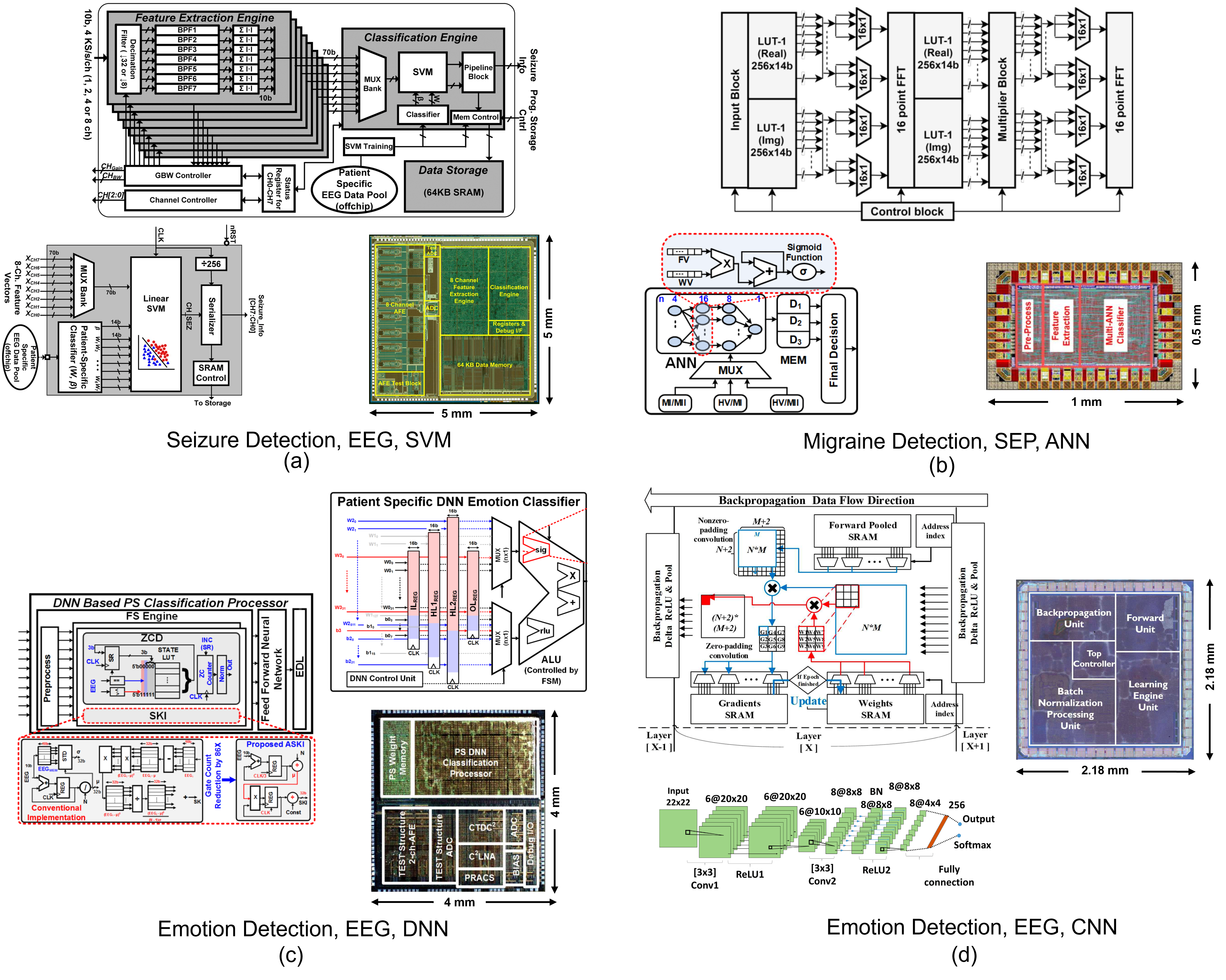}
	\vspace{-3mm}
	\caption{Hardware architectures and chip micrographs of ML-embedded neural prostheses for various applications: (a) Linear SVM for epilepsy \cite{yoo20128}, (b) ANN for migraine state detection \cite{taufique2021low}, (c) DNN for Autism emotion detection  \cite{aslam2020a10},  (d) CNN for emotion detection \cite{fang2019development}.}
	\label{figsoc2}\vspace{-3mm}
\end{figure*}

Most ML-embedded SoCs for epilepsy have adopted classifiers based on support vector machines (SVMs), as shown in \textcolor{black}{Fig.~\ref{figsoc1} and Fig.~\ref{figsoc2}(a).} 
Several variants of SVM kernels including linear, second-order polynomial, and radial basis function (RBF) have been reported for on-chip implementation. An SVM classifier generates weighted feature matrices using multiply-and-accumulate (MAC) blocks and separates them into different classes via linear or non-linear separation boundaries. 
For example, \cite{yoo20128} reported an 8-channel  linear SVM classifier  with  digital bandpower features implemented using a distributed quad-LUT  architecture, Fig. \ref{figsoc2}(a). The system was verified on the  MIT PhysioNet EEG database from the Children's Hospital Boston (CHB-MIT). This dataset includes 906 hours of recordings from 24 patients with epilepsy with $\sim$190 registered seizures, and is commonly used in EEG-based seizure detection SoCs (Table. \ref{comparison1}). Alternatively, the design in \cite{altaf20151} implemented a Gaussian basis function (GBF) SVM classifier to account for linearly non-separable seizure patterns, Fig. \ref{figsoc1}(b). A natural log operator was employed to linearize the GBF equation and replace multiplications with additions.  Time-division multiplexing was used to implement the bandpower features in an area- and energy-efficient manner. The non-linear SVM typically requires sufficient seizure patterns for training, which might be impractical for patients with limited training sets. Later, a combination of two linear SVMs was introduced \cite{altaf201516} to address this limitation, Fig. \ref{figsoc1}(a). The two SVMs were trained separately to achieve high sensitivity and specificity, and the classification results were combined to generate final decisions. This noninvasive closed-loop  SoC integrates a transcranial electrical stimulator (tES) to suppress impending seizures. The classification performance and ASIC specifications are summarized in Table. \ref{comparison1}. 


A 32-channel closed-loop neuromodulation system integrating frequency and phase-domain features, a 32-to-4 autoencoder for dimensionality reduction, and an exponentially decaying memory SVM (EDM-SVM) was proposed for seizure control  \cite{o2018nurip}, Fig. \ref{figsoc1}(f). This system was validated on 500 hours of  iEEG  data (4 patients, 44 seizures) provided by the EU dataset. 
The design proposed in \cite{wang202026} is an 8-channel closed-loop neuromodulation system for DBS, that was verified using stereo-EEG (sEEG) electrodes. The classifier is composed of a two-level coarse/fine detector, in which the DSP chip (separate from the core sensing chip) is only activated in case of suspected seizures raised by the coarse detector.
In this mode, maximum-modulus discrete wavelet transform (MODWT) and kernel density estimation (KDE) are computed and classified by a  least-squares SVM (LS-SVM) for fine classification, Fig. \ref{figsoc1}(h). Furthermore, \cite{lee2013low} reported a configurable SVM processor  with various kernels (RBF, polynomial, linear), validated  on the MIT EEG dataset.

It should be noted that in addition to the machine learning processor, the feature extraction circuits can be  highly power- and area-demanding, particularly in  systems with many input channels. Minimizing the number of extracted features and their hardware complexity without jeopardizing the classification accuracy is essential to reduce the overall energy consumption and area. 
The required computational resources in SVM linearly scale with the number of neural channels, making such optimizations more critical in practice.

An 8-channel closed-loop iEEG-based seizure control SoC was presented in \cite{chen2014fully}, computing  frequency spectrum and time-domain entropy  along with a linear least-square classifier, Fig. \ref{figsoc1}(c). This system was acutely verified in Long-Evans rats. Similarly, the closed-loop 16-channel design in \cite{cheng2018fully}  integrated a biosignal processor to extract approximate entropy (ApEn) and FFT-based bandpower features, passed to  a ridge regression classifier (RRC). The system was verified on ECoG data from five patients (duration not reported), and acutely for closed-loop seizure suppression in mini-pigs, Fig. \ref{figsoc1}(d). 



In addition to the above models, machine learning algorithms that exploit decision trees, either as base estimators in  ensemble methods  such as bagging and boosting \cite{shoaran2016hardware, shoaran2018energy, o202026} or as stand-alone classifiers \cite{zhu2020resot} have been used in neural signal classification tasks. 
While Random Forests \cite{breiman2001random} apply a bagging technique to DTs in order to reduce variance, boosting is a bias reduction technique in which individual trees are incrementally added to the ensemble to correct the previously misclassified samples. Popular implementations of boosting methods include  gradient boosting \cite{chen2016xgboost} and AdaBoost \cite{freund1997decision}. Both bagging and AdaBoost  use classifiers as base estimators, while  gradient boosting requires regressors. 
Particularly, ensembles of gradient-boosted DTs have recently emerged as an accurate \cite{kuhlmann2018epilepsyecosystem}, yet hardware-efficient \cite{shoaran2016hardware, shoaran2018energy, shoaran2017efficient} machine learning solution for neural SoC platforms and for applications with limited training sets. 
DT ensembles avoid hardware-intensive MAC operations and enable low-complexity hardware architectures for neural prosthesis applications. 


In \cite{shoaran2018energy}, a gradient-boosted DT ensemble achieved a record energy efficiency (41.2nJ/class, 32-channel) and a compact area (1 mm$^2$) for seizure detection, Fig. \ref{figsoc1}(e). The system was validated on iEEG  from 26 epilepsy patients (3074 hours, 393 seizures), available on the iEEG portal \cite{wagenaar2015collaborating}, a collaborative  platform for sharing large iEEG datasets.
An on-demand feature extraction approach was adopted  by sequentially using a single  feature extraction unit  in each tree, thus substantially reducing the number of extracted features and the overall hardware cost for inference. As opposed to other classifiers that compute all features for each input channel, this unique property of DTs allows the classifier to  selectively extract a limited number of features to minimize the loss function, thus accommodating a higher number of input channels   (Table. \ref{comparison1}). 
Another CMOS implementation of tree-based models used AdaBoost with 1024  trees of  depth one  for seizure detection and closed-loop stimulation \cite{o202026}, Fig. \ref{figsoc1}(g). 
Thanks to a bit-serial processing scheme, this 8-channel SoC reported state-of-the-art energy efficiency  (36nJ/class) for 8-channel iEEG classification.
Recent work  replaced axis-aligned splits with logistic regression to construct powerful oblique trees  as an efficient  combination of neural networks and DTs \cite{zhu2020resot} (Section VI) for epileptic seizure and PD tremor detection.





\begin{table}[t!]
\centering
\caption{Comparison of Machine Learning SoCs}
\vspace{-2mm}
\scalebox{0.72}{
\begin{tabular}{l|c|c|c}
\hline \hline
Parameter & TCAS-II'21 \cite{taufique2021low} &  JETCAS'19 \cite{fang2019development} &  CICC'20 \cite{aslam2020a10} \\
\hline \hline
Process & 180 nm & 28 nm & 180 nm\\
\hline
Classifier & Multi-ANN$^+$ & CNN & DNN\\
\hline
Application & Migraine Detection &  Emotion Detection & Emotion Detection \\
\hline
Features & HFO, BPF, Peak latency & Off-chip & ZCD, SK\\
\hline
Signal Modality & SEP & EEG & EEG \\
\hline
Closed-loop &  N & N & N\\
\hline
\# of Sensing Channels & 1 & 6 & 2\\
\hline
ML Energy Efficiency & N.A. & N.A. & 10.13 $\mu$J/class. \\
\hline
ML Power & 249 $\mu$W& 76.61 mW & N.A. \\
\hline
Total Area (ML Area) & 0.5 (0.5) mm$^2$  & 3.35  (3.35) mm$^2$ & 16  (6.02$^\ast$) mm$^2$ \\
\hline
Sampling Rate/Ch. & 5 kS/s & 250 S/s & N.A. \\
\hline
Accuracy & 76\% & 83.4\%$^\ast\ast$ & 85.2\%\\
\hline
Dataset (\# patients) & MI, MII (42), HV (15) & DEAP (32) & DEAP (32), SEED\\
\hline
Latency & 50 ms$^\dag$ & 0.45 s$^\dag$ & $<$1min$^\dag$ \\
\hline
ML Energy/Ch. & 49.8 nJ/S & 51.1 $\mu$J/S & N.A. \\
\hline
ML Area/Ch. & 0.5 mm$^2$ & 0.558 mm$^2$ & 3.01 mm$^2$ \\
\hline
ML E-A FoM & 24.9 nJ$\cdot$mm$^2$/S & 29 $\mu$J$\cdot$mm$^2$/S & N.A. \\
\hline \hline
\multicolumn{3}{l}{$^+$ Post place-and-route results.} & \\
\multicolumn{3}{l}{$\ast$ ML (feature extractor and classifier) area estimated from chip micrograph} \\
\multicolumn{3}{l}{$\ast\ast$ Accuracy metric} & \\
\multicolumn{3}{l}{$\dag$ Processing (system) latency} \\
\end{tabular} 
}
\label{comparison2} \vspace{-3mm}
\end{table}

\vspace{-1mm}
\subsection{Implants for Movement Disorders}\vspace{-1mm}
Multiple feasibility studies using closed-loop DBS devices like Medtronic's Percept and Summit have demonstrated additional benefits using closed-loop versus open-loop DBS in movement disorders \cite{swann2018adaptive, shute2016thalamocortical}.
Closed-loop DBS in PD \cite{little2013adaptive, meidahl2017adaptive, arlotti2016adaptive} has led to improvements in tremor control, reduced stimulation time and power consumption, and reduced speech side effects compared to open-loop DBS. However, wider adoption of this approach is awaiting advances in implantable hardware, control algorithms, and chronic validation. 
Current systems  predominantly use single-biomarker thresholding, which precludes the optimized control of tremor. 

Recently, ML approaches have been used for detecting motor symptoms (e.g., tremor)  in patients with PD and essential tremor \cite{wang2018towards,yao2020improved, yao2018resting, zhu2019cost, he2021closed, watts2020machine} to control DBS in closed loop. 
An approach based on feature engineering and tree boosting \cite{yao2020improved, yao2018resting} used various correlating features of tremor such as bandpower in multiple frequency bands, the ratio of  high-frequency oscillations, phase-amplitude coupling, and tremor power to detect the onset of rest-state tremor episodes in PD. Using only five selected features, the system was able to predict tremor with a 89.2\%  sensitivity  and detection lead of 0.52s in 12 patients, significantly better than conventional beta-thresholding approach. Fixed-point quantization and power-aware inference were later used to enable low-power gradient boosting, achieving  55.4\% energy reduction compared to conventional tree ensemble \cite{zhu2019cost}. A method based on resource-efficient oblique trees (ResOT) was recently applied to PD tremor detection, enabling significant energy and memory reduction by various hardware-algorithm co-design techniques \cite{zhu2020resot}.
A similar study was recently done on patients with  essential tremor (ET) \cite{he2021closed} who suffer from tremor during voluntary movements. Using a binary classifier, postural tremor and voluntary movements  were detected from LFP  features recorded via the DBS lead, achieving an average sensitivity of 80\% in 7 patients with ET.
Such machine learning techniques hold the promise to enable accurate symptom detection in closed-loop neural prostheses for various movement disorders. More developments in  SoC design for such applications are expected in near future.


\vspace{-1mm}
\subsection{Implants for Neuropsychiatric Disorders and Memory} \vspace{-1mm}
Neuromodulation, particularly invasive technologies like DBS, has  been recently explored for treating psychiatric disorders such as major depressive disorder (MDD) and obsessive compulsive disorder (OCD) \cite{lozano2012multicenter, krauss2020technology}. However, despite promising early results, the high-profile clinical trials have shown inconsistent effects. One major limiting factor is the open-loop approach used in conventional DBS, which has been shown to be inefficient in  engaging target brain regions  in complex disorders such as depression and OCD  \cite{lo2017closed}. While the application of neurostimulation techniques has made a significant impact on the lives of patients with movement disorders, major advances are needed to treat more prevalent conditions such as depression. Closed-loop patient-specific stimulation appears to be the most viable solution. 

Development of algorithms for  automated detection of emotional states and shifts in arousal, vigilance, and wakefulness has received considerable attention in EEG-based human studies, with some recent reports on SoC design. For example, a  deep neural network (DNN) classifier  was implemented   for emotion detection in  children with Autism \cite{aslam2020a10}. The valence/arousal binary classification by the 4-layer DNN was used to detect four-state emotions. A reduction in energy consumption was achieved through a pipelined DNN architecture with a central arithmetic logic unit, Fig. \ref{figsoc2}(c). This DNN processor can analyze two EEG channels with an accuracy of 85.2\% and energy efficiency of 10.1 $\mu$J/class.  In another design, a convolutional neural network (CNN)  was proposed for emotion detection  \cite{fang2019development}, offering an online training feature, Fig.~\ref{figsoc2}(d). To minimize   area and memory overhead due to batch processing,   hardware re-use and minibatch data were employed for training and acceleration, at the expense of longer training time. Using an external feature extraction engine, this system obtained a 83.36\% accuracy in binary  classification of emotions (Table. \ref{comparison2}). Machine learning has also been  explored in sleep stage classification \cite{chang2019ultra}, task engagement \cite{provenza2019decoding} and mental fatigue prediction~\cite{yao2020mental} to potentially trigger a neurostimulation therapy. 




Disorders such as Alzheimer’s disease exhibit network abnormalities, necessitating the need for multi-site electrophysiological recordings. The closed-loop stimulation approach in \cite{ezzyat2018closed} used a patient-specific logistic regression classifier to decode the brain-wide electrocorticography (ECoG) signals, and subsequently triggered stimulation in response to the predicted periods of poor memory encoding to enhance memory. 
The results suggest a predictive role of increased high-frequency as well as decreased low-frequency activity for memory recall, and that responsive neuromodulation in the lateral temporal cortex could improve recall performance. More developments in  neural prosthesis design for mental and memory disorders are expected in the coming years.

\vspace{-1mm}
\subsection{Wearables for Migraine} \vspace{-1mm}
While most current devices have been developed for epilepsy and movement disorders, there is an increasing demand for novel  therapeutic devices for other  medication-resistant neurological disorders. Migraine, for instance, is the most common neurological disorder that affects millions around the world.
Migraine patients suffer from episodic headaches lasting hours to days and often move from a stage of low-frequency attacks into chronic migraine. The diagnosis mainly relies on  patient diaries and clinical interviews \cite{restuccia2013different, zhu2019migraine}. As an emerging alternative, neurophysiological monitoring techniques have shown to be beneficial in assessing migraine progression \cite{restuccia2013different}. The automated  detection of migraine state using continuous brain recordings could help in early and more effective treatment, either with medications or neurostimulation. 

A machine learning approach was recently proposed  for noninvasive  migraine state detection from somatosensory evoked potential (SEP) biomarkers in 42 migraine patients, as described in \cite{zhu2019migraine}.  The results suggest the potential use of SEP as a feedback  signal for migraine attack prediction. Based on this  idea, \cite{taufique2021low} reported a low-power feature extraction and ML processor  for migraine state prediction, using single-channel SEP as input. Multiple features such as bandpower, time-domain and statistical features of high-frequency oscillations \cite{zhu2019migraine} were integrated with a multi-class artificial neural network (ANN), achieving a predictive accuracy of 76\%, Fig.~\ref{figsoc2}(b) (\textcolor{black}{chip layout post place-and-route}).

\vspace{-1mm}
\subsection{Implants for Stroke and Traumatic Brain Injury}  \vspace{-1mm}
Neurostimulation can be used to facilitate post-stroke plasticity and functional recovery. 
Compared to noninvasive methods such as transcranial magnetic  or direct-current stimulation (TMS, tDCS), invasive tools such as direct cortical stimulation  offer a higher temporal and spatial resolution.
However, current  cortical stimulation approaches for stroke are limited by the poor localization of stimulation targets and open-loop operation \cite{plow2014invasive}, 
urging the need for advanced data analysis and machine learning techniques.

In addition, patients with severe-to-moderate traumatic brain injury (smTBI) suffer from persistent cognitive dysfunction and chronic mental fatigue that significantly impacts all aspects of their functioning. Despite extensive efforts to develop rehabilitation and medication-based therapies, there are no effective therapeutic options for these patients. In a breakthrough study, it was shown that therapeutic DBS in the central thalamus  (CT-DBS) could restore  executive function, fluent communication and motor control in a patient who remained  in a  minimally conscious state for six years following a TBI \cite{schiff2007behavioural}. Similar improvements have been observed in individuals with chronic mental fatigue.
In a recent study, the  ECoG activity from two healthy non-human primates (NHPs) during a sustained attention task was used to predict the onset of mental fatigue \cite{yao2021predicting, yao2020mental}. Using spectrotemporal and connectivity biomarkers and a tree ensemble classifier, the decline in animal's performance was predicted,   seconds prior to NHP's behavioral response. 
This  approach could potentially be used for closed-loop neurostimulation in patients with TBI.

In a proof-of-concept study \cite{guggenmos2013restoration}, a closed-loop neural  SoC was used to facilitate recovery after brain injury in a rat model of  brain injury. The action potentials detected in premotor cortex were used to trigger neurostimulation in somatosensory cortex for several weeks. This spike-triggered stimulation led to significantly improved reaching and grasping functions, enhancing the functional connectivity between the two brain regions. These findings motivate the design of novel closed-loop neural prostheses to  treat brain injury and similar neurological indications.


\vspace{-2mm}
\subsection{Comparison of ML-embedded SoCs} 
A comparison on  hardware specifications and classification performance of state-of-the-art neural prostheses with on-chip machine learning is presented in Table~\ref{comparison1} (for epilepsy) and Table~\ref{comparison2} (for other applications). When comparing the  performance and hardware cost of different ML SoCs, one should consider various factors that affect the overall predictive performance and design complexity, such as the  input signal modality and dataset, the number of analyzed patients, the duration of recording and seizure count, and the metrics used to evaluate the algorithm/hardware performance (e.g., accuracy, F1 score,  sensitivity, power vs. energy efficiency, detection vs. system latency). In addition, the number of processed channels  should be taken into account to fairly compare various architectures and assess their scalability. 

\textcolor{black}{Energy efficiency has been a common metric to compare different ML-embedded biomedical SoCs in literature. However, we note that the energy efficiency is not being reported in a unified manner (e.g., total power consumption/sampling rate \cite{yoo20128, shoaran2018energy, altaf201516, altaf20151, o202026} or total power consumption/classification rate \cite{o2018nurip, chen2014fully, cheng2018fully} has been used), which may hinder appropriate design choices. Furthermore, the number of channels is not taken into account, which is particularly important in modern neural prostheses. Here, we define a new energy-area efficiency figure of merit (E-A FoM) as follows:
\begin{equation}
\text{E-A FoM} = \frac{{P_{Ch}}\cdot{A_{Ch}}}{f_{s}}
\end{equation}
where $P_{Ch}$ and $A_{Ch}$ indicate per-channel power and area of the ML SoC, respectively, and $f_{s}$ is the per-channel sampling rate of the signal processing circuits. Similar FoMs have been used in AFE and ADC design for multi-channel neural recording \cite{park2017modular}. The E-A FoM fairly represents the energy-area efficiency of the system while also factoring in the multi-channel scalability. Other performance metrics such as accuracy and latency are excluded as those metrics can vary among different datasets and applications. Tables \ref{comparison1} and \ref{comparison2} report the E-A FoM of the state-of-the-art neural prostheses along with their per-channel area and energy consumption. Only the power and area of the ML processor (i.e., feature extractor, classifier, and memory for parameter storage) have been considered. This FoM indicates that the tree-based models achieve  orders of magnitude superior energy-area efficiency compared to SVM classifiers, while providing comparable classification accuracy and latency. With cost-aware hardware-algorithm co-design, we aim to improve the efficiency of tree-based classifiers even further, as discussed in Sections VI-VII.}

\textcolor{black}{The predictive power and hardware efficiency of different SoCs are greatly affected by their selection of ML algorithms. For example, DT-based ML models feature a lightweight inference scheme where we simply compare feature values to thresholds to route samples through the tree. On the contrary, the inference of kernelized SVM involves vector multiplications and the calculation of Gram matrix, which partially explains the E-A superiority of DTs over SVMs in Table \ref{comparison1}. Moreover, inspired by the recent success of deep learning algorithms, there is an increasing interest in deploying CNNs and DNNs on neural SoCs \cite{fang2019development,taufique2021low,aslam2020a10}. However, compared to conventional approaches, deep learning models generally require more training data and consume higher power consumption \cite{taghavi2019hardware}. The benefits of using deep learning in neural SoCs need to be further investigated in the future.}

\vspace{-2mm}
\subsection{\textcolor{black}{Limitations of the Current SoCs and Future Directions}} \vspace{-1mm}

\textcolor{black}{High-density electrode arrays have shown promise in both neurophysiological monitoring \cite{hermiz2018sub} and therapeutic neurostimulation \cite{mandal202146}. However, the channel count of state-of-the-art ML SoCs is limited to 32, which could hinder their clinical application. The most critical challenges to realizing high-channel-count ML-embedded neural prostheses lie in the AFE, the back-end signal processing, and the memory for parameter storage. Over the past years, the field has witnessed a growth of channel count in neural prostheses, such as Neuralink's BMI platform with 3072 channels \cite{musk2019integrated}. Recently, a 1024-channel closed-loop BMI SoC was presented with a successful demonstration of motor intention decoding (performed offline) in a macaque monkey \cite{yoon20211024}. Novel area- and power-efficient AFE design techniques (such as mixed-signal \cite{muller2014minimally} and time-division multiplexing  \cite{sharma2018acquisition, uehlin20190}) should continue to be explored. This will enable advanced neural prostheses with high resolution, reduced invasiveness, longer lifetime, and minimized heat-induced tissue damage. In addition to area-power constraints on the AFE, the burden of the back-end signal processing (i.e., feature extraction and classification) is a major bottleneck to next-generation high-channel-count prostheses. The amount of computation in the current ML SoCs grows linearly with channel count, posing a major challenge on the energy consumption. The on-demand feature computation scheme in \cite{shoaran2018energy} could be a viable  solution to realizing a scalable ML SoC. Only relevant features from a subset of channels are computed in each processing window, achieving a substantial reduction in hardware cost. Similar techniques will pave the way for the integration of  novel high-density  electrodes (Section II. A) in future diagnostic and closed-loop devices.
Another on-demand processing approach was adopted in an SVM-based two-level (coarse/fine) classifier to reduce the system power consumption \cite{wang202026}. Exploiting the sparseness of seizures, the otherwise power-demanding SVM  classifier (fine) in a separate chip is only activated upon seizure declaration by the coarse detector. The two-level SVM classifier performs 266 classifications/hour with 1.16 $\mu$W average power, improving $>$135$\times$ over the conventional SVM. Single-chip integration and multi-channel scalability have yet to be addressed with this approach.}

\textcolor{black}{In addition, most current  classifiers integrated on neural prostheses  use an offline training scheme with fixed  parameters, thus neglecting the  non-stationary dynamics of neural signals. The next generation ML-embedded neural SoCs are expected to perform active, incremental learning to account for the previously unseen changes in neurological patterns. In online machine learning, the model parameters are updated with the sequential arrival of data, thus dynamically adapting to new signal patterns. Online learning algorithms have shown promise in stable chronic neural decoding \cite{wang2013online}. Yet, the deployment of such models on ASIC with minimal area and power consumption remains an open direction. Current on-chip systems based on SVM \cite{huang20191, feng2017vlsi} are highly energy and memory demanding, while off-chip recalibrations pose security risks and reduce patient independence.  More developments in this area are expected in near future.}

The ML-embedded neural prostheses, like other edge AI devices in  IoT and healthcare, may greatly benefit from developments in algorithm and circuit design that could lead to higher performance, lower energy and more compact area. For instance, future ML SoCs are expected to benefit from emerging techniques in CMOS design such as analog, mixed-signal \cite{murmann2015mixed}, and approximate computing, 
as well as in-memory computing techniques.
Particularly, in-memory computing has shown the potential to achieve remarkable improvements over  conventional digital implementations \cite{zhang2017memory, rahimiazghadi2020hardware, chen2015128}. \textcolor{black}{Compared to current SoCs, neuromorphic hardware integrates spiking neural networks (SNN) and in-memory computing to avoid the communication overhead between processors and memory, and allows unsupervised online learning via Spike Timing Dependent Plasticity (STDP). As discussed in \cite{rahimiazghadi2020hardware}, currently the memristor-based designs are rarely used  in the biomedical domain. Moreover, it should be noted that the decoding performance of SNN is relatively low 
due to the lack of maturity of the training algorithms \cite{rahimiazghadi2020hardware}. Deploying high performance SNN and memristor-based designs  in neural prostheses remains as a future direction. }

\vspace{-1mm}
\section{Hardware-Algorithm Co-Design of \\Decision Tree Ensembles} \vspace{-1mm}
Designing  machine learning models that consume little energy and area, while  providing a  high classification accuracy and low detection latency is essential to the next-generation smart neural prostheses. As discussed in Section IV, decision trees are widely used in edge applications and  neural decoding tasks thanks to their low inference complexity, easy and fast training, as well as high predictive power in ensemble methods or oblique structures \cite{shoaran2018energy, zhu2020resot, o202026, zhu2020closed, kuhlmann2018epilepsyecosystem}.
These advantages are essential for extremely resource-constrained  platforms such as a brain implant or wearable with high channel counts. 
In this \textcolor{black}{section}, we present novel approaches  to optimize the key  design metrics of an on-chip DT ensemble, including the power consumption and processing latency, in the context of neural signal classification tasks. Some of these techniques are broadly applicable to other machine learning algorithms for various  implantable and edge applications.
\vspace{-1mm}
\subsection{Depth-Variant Tree Ensemble for Latency Reduction} \vspace{-1mm}
In a decision tree, test sample traverses a single root-to-leaf path during inference \cite{zhu2020resot, tanno2018adaptive}. Despite being lightweight and area-efficient, the single-path scheme requires conditional computation and evaluates nodes in a sequential order \cite{hazimeh2020tree}. As a result, DT-based classifiers impose a latency that  increases proportionally with the decision path length. 
However, early symptom detection is critical to effectively treat neurological disorders, and it is directly affected by the processing latency. Previous work reduced seizure detection latency  by either using shorter windows \cite{burrello2020ensemble} or replacing the widely used bandpower biomarkers with new features such as neuronal potential similarity \cite{bandarabadi2015early}.
However, such methods may suffer from a degraded classification performance (since low-frequency features that require a longer window could be  critical in symptom detection \cite{burrello2020ensemble}) or poor generalizability due to the use of specific biomarkers \cite{bandarabadi2015early}. 
To the best of our knowledge, this study is the first  to  address latency reduction from an algorithmic perspective.

\begin{figure}[t!]
  \centering
  \vspace{0mm}
  \includegraphics[width=1\columnwidth]{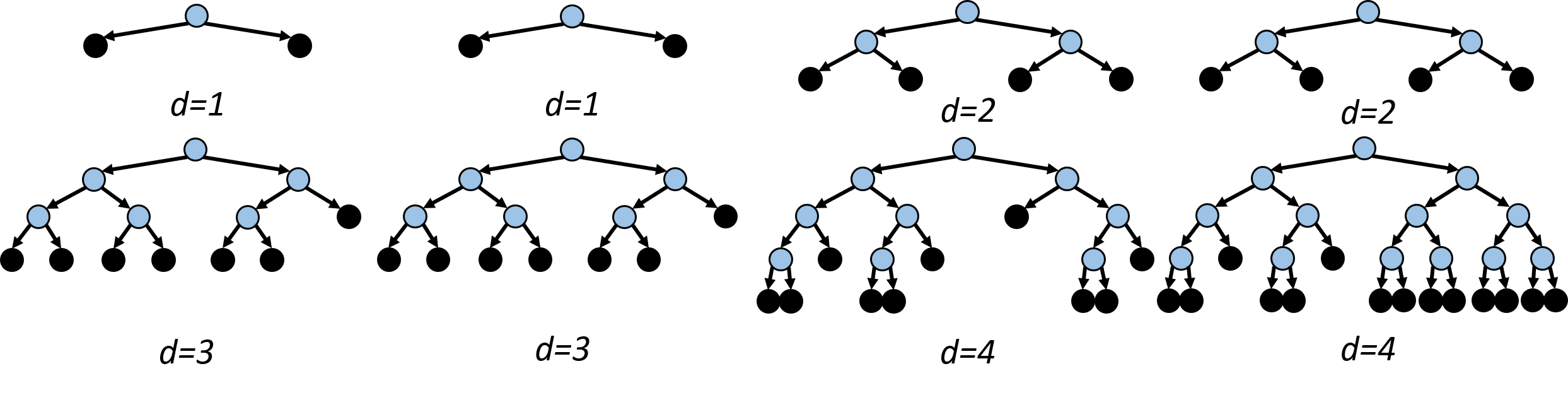}
  \vspace{-6mm}
  \caption{A DVTE with eight decision trees. Unlike conventional tree ensembles that uniformly set the maximum depth on all trees, the maximum depths in a DVTE are different (1--4). The internal and leaf nodes  are shown in blue and black, respectively.}\vspace{-2mm}
  \label{f1}
\end{figure}

Tree ensembles have shown promise in various neural classification tasks \cite{shoaran2018energy, kuhlmann2018epilepsyecosystem, yao2020improved, zhu2020resot}.
However, conventional ensembles impose a uniform maximum-depth constraint on all base-estimators in the ensemble, such that the system latency is similar across different trees. In this work, we propose the Depth-Variant Tree Ensemble (DVTE),  a novel low-latency variation of conventional ensemble methods. As shown in Fig.~\ref{f1}, DVTE consists of decision trees with different maximum depths, resulting in non-uniform  latencies across trees. In a DVTE, shallow trees perform fast inference to reduce system latency, while deep trees are trained to compensate for misclassification errors caused by shallow trees.

\begin{figure}[t!]
  \centering
  \includegraphics[width=1\columnwidth]{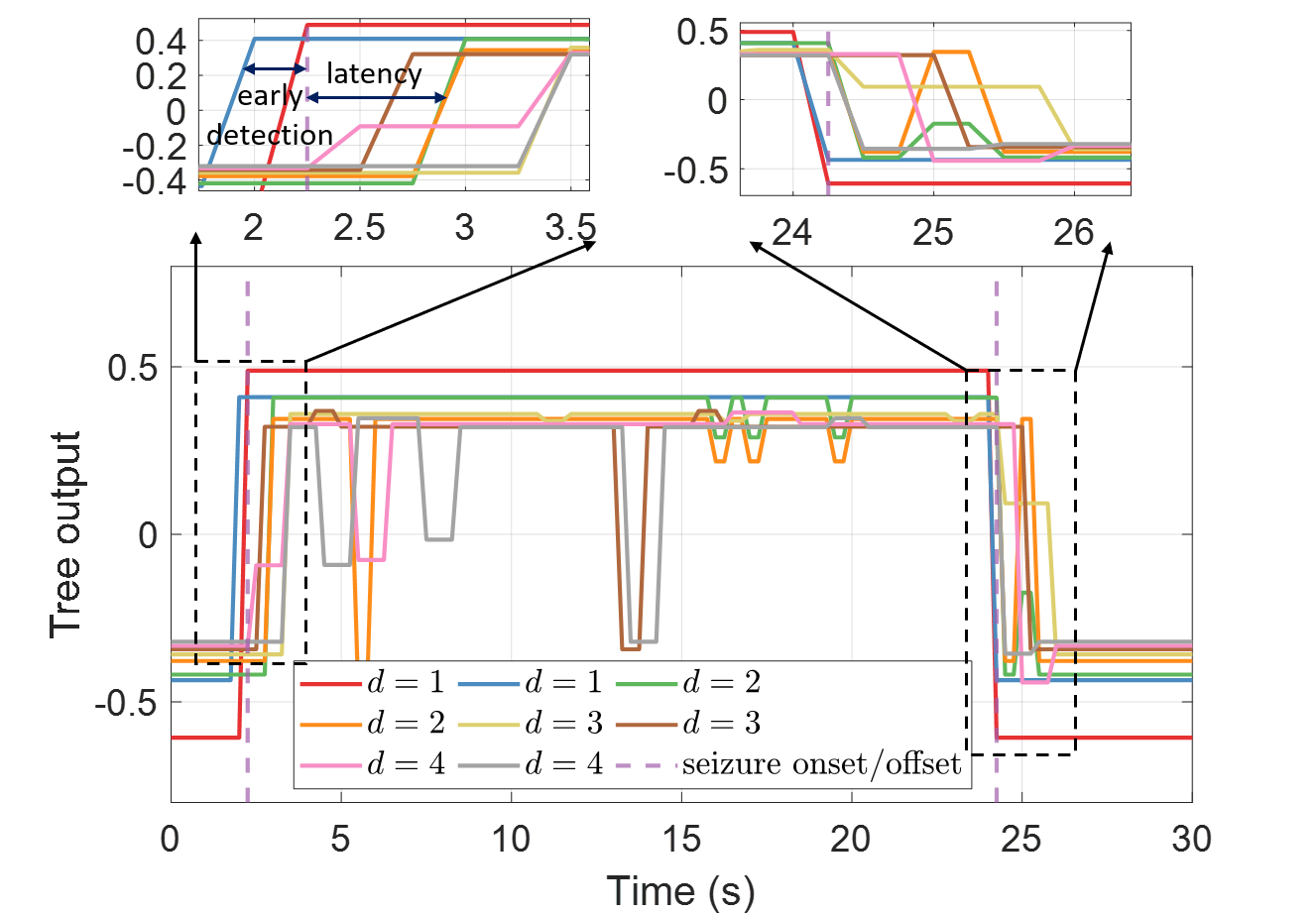}
  \vspace{-5mm}
  \caption{The outputs of decision trees in a DVTE. Latency is defined as the time difference between the expert-marked seizure onset  and the state change of each tree's output. 
  $d$ is the maximum depth of each tree.}\vspace{-2mm}
  \label{f2}
\end{figure}

We trained the proposed DVTE model using the popular gradient boosting framework \cite{chen2016xgboost,ke2017lightgbm}. In the first two boosting rounds, we initialized the ensemble with decision stumps (i.e., decision trees with a single internal node). In the third and fourth rounds, two DTs with a max depth of two were trained to compensate for the residual errors from previous rounds. Deeper  trees were gradually added to DVTE in later boosting rounds to better fit on training data. During inference, all decision trees in a DVTE run freely in parallel, with no need for synchronization. 
Therefore, shallow trees can update the decision outcome more frequently than deeper trees.  If the current inference in a deep tree is incomplete (i.e., test samples have not yet reached the leaf nodes), we used the most recent output of that tree. The final prediction of DVTE is calculated as the sum of the outputs of all trees in the ensemble, which can be updated at the same rate as the shortest tree (i.e., $d = 1$).
While shallow trees make predictions with low latency (trees of $d=1$ in Fig. \ref{f2}), they often have a limited predictive power and may not fit well on training data. To tackle this problem and achieve the best trade-off between latency and classification accuracy, DVTE incorporates deeper trees in the gradient boosting framework to reduce bias.

Unlike DVTE which effectively combines shallow and deep trees in the gradient boosting ensemble to jointly optimize the latency and accuracy, previous work either used a few deep trees (e.g., 8 trees with a max depth of 4  \cite{shoaran2018energy}) with potential   latency concerns as discussed above, or implemented a large number of shallow trees (1024 decision stumps in \cite{o202026}), requiring many parallel feature processing units. The aim of DVTE is to benefit from both shallow and deep trees  and enable low-latency inference with  a small tree ensemble. This is particularly critical in time-sensitive classification tasks such as PD tremor detection with strict latency requirements.

As an example, we built a DVTE with 8 trees and various depths from 1 to 4 (Fig. \ref{f1}). This model was benchmarked against conventional ensemble  (8 trees, max depth: 4 \cite{shoaran2018energy}). We used a learning rate of 0.3 for both models and implemented them using the lightGBM library in Python \cite{ke2017lightgbm}.  We tested our classifier on  epileptic seizure detection  using iEEG recordings (11 patients, 106 annotated seizures over 1255 hours). The number of channels varied from 47 to 128. This dataset can be accessed via  iEEG portal \cite{wagenaar2015collaborating}. Handcrafted features were extracted over various window lengths as detailed in Table~\ref{tab1}. \textcolor{black}{It should be noted that both EEG and iEEG have been widely used in on-chip seizure detectors \cite{shoaran2018energy,chen2014fully,o202026, cheng2018fully, o2018nurip, wang202026, altaf20151, yoo20128,altaf201516}. 
However, iEEG is more commonly used in closed-loop prostheses, as it can be easily combined with invasive neuromodulation techniques for improved symptom control \cite{chen2014fully,o202026, cheng2018fully, o2018nurip}, and it has been used in our study.
}

\begin{figure}[t!]
  \centering
  \includegraphics[width=1\columnwidth]{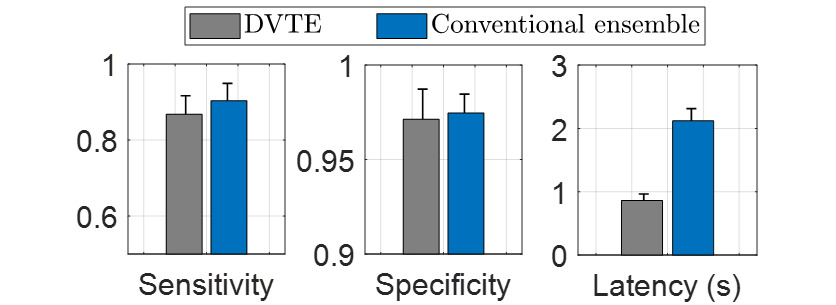}
  \vspace{-4mm}
  \caption{Performance comparison of DVTE and conventional tree ensemble with a maximum depth of 4. DVTE reduced the latency by 2.5$\times$ with a marginal performance reduction ($<$3\% in sensitivity and $<$1\% in specificity). Error bars indicate the standard errors across patients.}\vspace{-3mm}
  \label{f3}
\end{figure}

Figure \ref{f3} compares the proposed DVTE and the conventional ensemble method in terms of classification performance (sensitivity, specificity) and latency. \textcolor{black}{The performance was evaluated using bit-accurate classifier models in MATLAB and Python.} We estimated the processing latency by calculating the average time to traverse a root-to-leaf decision path in the trees. Compared to the conventional ensemble, DVTE caused a marginal  performance reduction ($<$3\% in sensitivity and $<$1\% in specificity). On the other hand, DVTE achieved an average latency of 0.86s,  significantly lower than the latency of a conventional ensemble (2.12s, 2.5$\times$ reduction). 

\vspace{-2mm}
\subsection{Cost-Aware Learning for Latency and Power Reduction} 
The inference phase of tree-based models is relatively simple. In axis-aligned decision trees, we compare a feature value to a threshold in order to select the child node at each internal node. The leaf node contains a constant weight indicating the prediction result. 
Given the lightweight inference in tree-based models, the 
hardware cost (e.g., power, latency) is largely affected by the feature extraction process~\cite{shoaran2018energy}. 

\textcolor{black}{Table~\ref{tab1} summarizes the biomarkers used in our seizure detection task and their power and latency cost. We implemented digital feature extraction hardware in a TSMC 65 nm LP process using Synopsys Design Compiler and Cadence Innovus. The power cost of each feature  was simulated under a 1.2-V supply using Synopsys PrimeTime.} Line-length, a widely used feature in epilepsy studies, is hardware-friendly and low-power. Bandpower features, on the other hand, consume higher power since they require an FIR filtering stage. The latency associated with a  feature depends on the window size used to compute that  feature. 
Long windows were used to extract low-frequency bandpower, while short windows were used for time-domain features and high-frequency bandpowers. Specifically, we used 1s windows to extract Delta ($\delta$), 0.5s for Theta ($\theta$) and Alpha ($\alpha$), and 0.25s  for other features. 

\begin{figure}[t!]
  \centering
  \vspace{-3mm}
  \includegraphics[width=0.7\columnwidth]{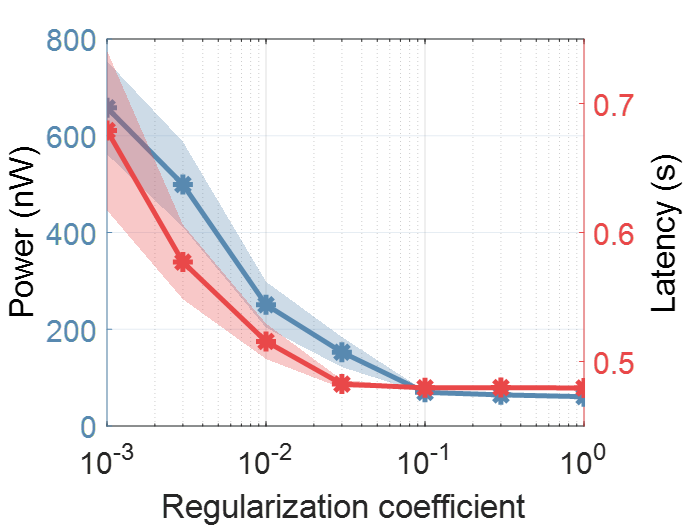}
  \vspace{-1mm}
  \caption{Hardware cost as a function of the regularization coefficient $C$ in DVTE. Large $C$ imposes strong regularization and reduces the power/latency cost. 
  The power cost was calculated as the average power consumption to extract features along the decision path. Latency was estimated as the average time to traverse a root-to-leaf decision path in the tree.}\vspace{-3mm}
  \label{f5}
\end{figure}

We apply the cost-aware learning approach to  tree-based classifiers (e.g., DVTE) to reduce the inference hardware cost. Specifically, we use the total power consumption and latency along the decision path as a regularization term  in the training process. The training of cost-aware decision trees attempts to minimize the following expression: 
\begin{equation}
\min  \sum_{i} L\left(y_{i}, f\left(\boldsymbol{x}_{i}\right)\right)+ C (\Psi_{pow}\left(f, \boldsymbol{x}_{\boldsymbol{i}}\right) + \Psi_{lat}\left(f, \boldsymbol{x}_{\boldsymbol{i}}\right)),
\label{eq1}
\end{equation}
where $L\left(y_{i}, f\left(\boldsymbol{x}_{i}\right)\right)$ is the loss function that measures the misclassification error as the difference between  groundtruth $y_{i}$ and prediction $f\left(\boldsymbol{x}_{i}\right)$, $\Psi_{pow}$ and $\Psi_{lat}$ indicate the estimated power consumption and latency along the decision path, respectively, and $C$ is the regularization coefficient that determines the trade-off between hardware cost and performance. The effect of varying $C$ on latency and power in DVTE  is shown in Fig. \ref{f5}. For a greater regularization coefficient, cost-aware decision trees achieve a lower hardware cost. Since power and latency span over different ranges, we standardized the cost by removing the mean value and normalizing both power and latency to their unit variance.

\begin{table}[h]
\vspace{-3mm}
\caption{Epilepsy features, their
power and latency costs}
\begin{center}
\vspace{-3mm}
\scalebox{0.76}{
\begin{tabular}{rl|c|c}
\hline \hline
\multicolumn{2}{c|}{\textbf{Features and description}} &  \textcolor{black}{\textbf{Power (nW)}} & \textbf{Latency (s)}\\
\hline
Delta ($\delta$):& Bandpower over 1-4Hz &  \textcolor{black}{250.6} & 1 \\
Theta ($\theta$):& Bandpower over 4-8Hz &  \textcolor{black}{250.6} & 0.5 \\
Alpha ($\alpha$):& Bandpower over 8-13Hz &  \textcolor{black}{250.6} & 0.5 \\
Beta ($\beta$):& Bandpower over 13-30Hz &  \textcolor{black}{250.6} & 0.25 \\
Low-Gamma ($\gamma_1$):&  Bandpower over 30-50Hz &  \textcolor{black}{250.6} & 0.25 \\
Gamma ($\gamma_2$):& Bandpower over 50-80Hz &  \textcolor{black}{250.6} & 0.25 \\
High-Gamma ($\gamma_3$):&  Bandpower over 80-150Hz &  \textcolor{black}{250.6} & 0.25 \\
Ripple:&  Bandpower over 150-250Hz &  \textcolor{black}{250.6} & 0.25 \\
Line-Length (LLN):& \hspace{-2mm} $\frac{1}{d} \sum_{d}|x[n]-x[n-1]|$ $d=$window size \hspace{-2mm} &  \textcolor{black}{7.4} & 0.25 \\
Variance (Var):& \hspace{-2mm} $\frac{1}{d} \sum_{d}(x[n]-\mu)^{2},$ $ \mu=\frac{1}{d} \sum_{d}(x[n])$& \textcolor{black}{21.6}& 0.25\\
\hline \hline
\end{tabular}
}
\label{tab1}
\end{center}
\vspace{-3mm}
\end{table}

\begin{figure}[t!]
  \centering
  \includegraphics[width=1\columnwidth]{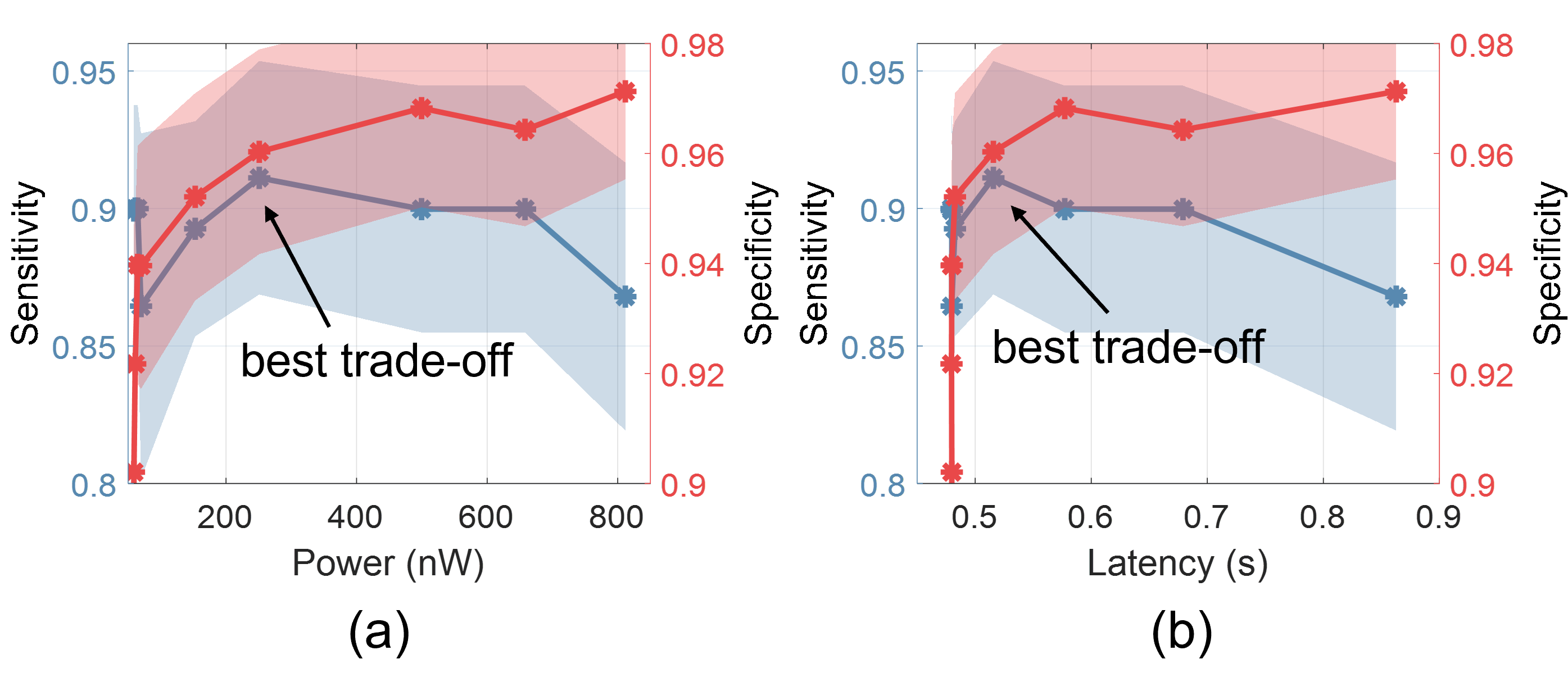}
  \vspace{-6mm}
  \caption{Seizure detection performance as a function of (a) power consumption and (b) latency. Shaded area indicates the standard errors across  patients. The experiment was performed using DVTE and the following setting: 8 trees, depths varying from 1 to 4.}\vspace{-4mm}
  \label{f6}
\end{figure}

We applied the cost-aware inference approach to DVTE to reduce both power and latency on  seizure detection task. Figure \ref{f6} shows the classification performance (sensitivity, specificity) as a function of the cost metrics (latency, power). We adjusted the regularization coefficient $C$ to achieve different trade-offs between power/latency and performance. For both the low-power (Fig.~\ref{f6}(a)) and low-latency (Fig.~\ref{f6}(b)) DVTEs, the best trade-off is observed at a point where a maximum reduction in latency or power can be achieved with only a marginal performance loss.

\begin{figure}[b!]
  \centering
  \includegraphics[width=1\columnwidth]{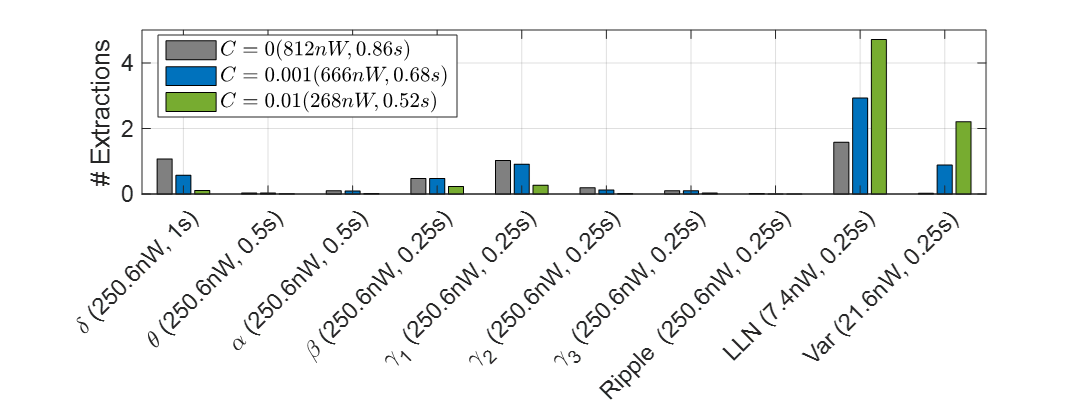}
  \vspace{-6mm}
  \caption{The number of extracted features in DVTE for different regularization coefficients. 
  With greater $C$, the cost-aware model tends to use hardware-friendly features (e.g., LLN, Var). Features with longer windows ($\delta, \theta, \alpha$) are also penalized in the cost-aware model. The power cost and latency  for each $C$  are shown in the legend, while the X-axis shows individual feature costs. For $C=0.01$, we achieved an average power cost of \textcolor{black}{268nW} and  latency of 0.52s. }
  \label{f4}
\end{figure}

Figure \ref{f4} shows the number of extracted features for the cost-aware DVTE. 
The number of feature extractions are normalized to each 0.25s  window. 
Thus, the normalized feature count is upper bounded by the number of trees. For $C>0$, we used the hardware cost to regularize the model and  as a result, DVTE was trained to minimize the inference power  and latency. As the regularization coefficient increases, the model further penalizes inefficient features. With $C=0.01$, we achieved the best trade-off between  performance and hardware cost (Fig. \ref{f6}), reducing the power by 3$\times$ and latency by 1.7$\times$ compared to DVTE without cost-aware learning.

\begin{figure*}[h] 
	\centering
	\includegraphics[width=1.8\columnwidth]{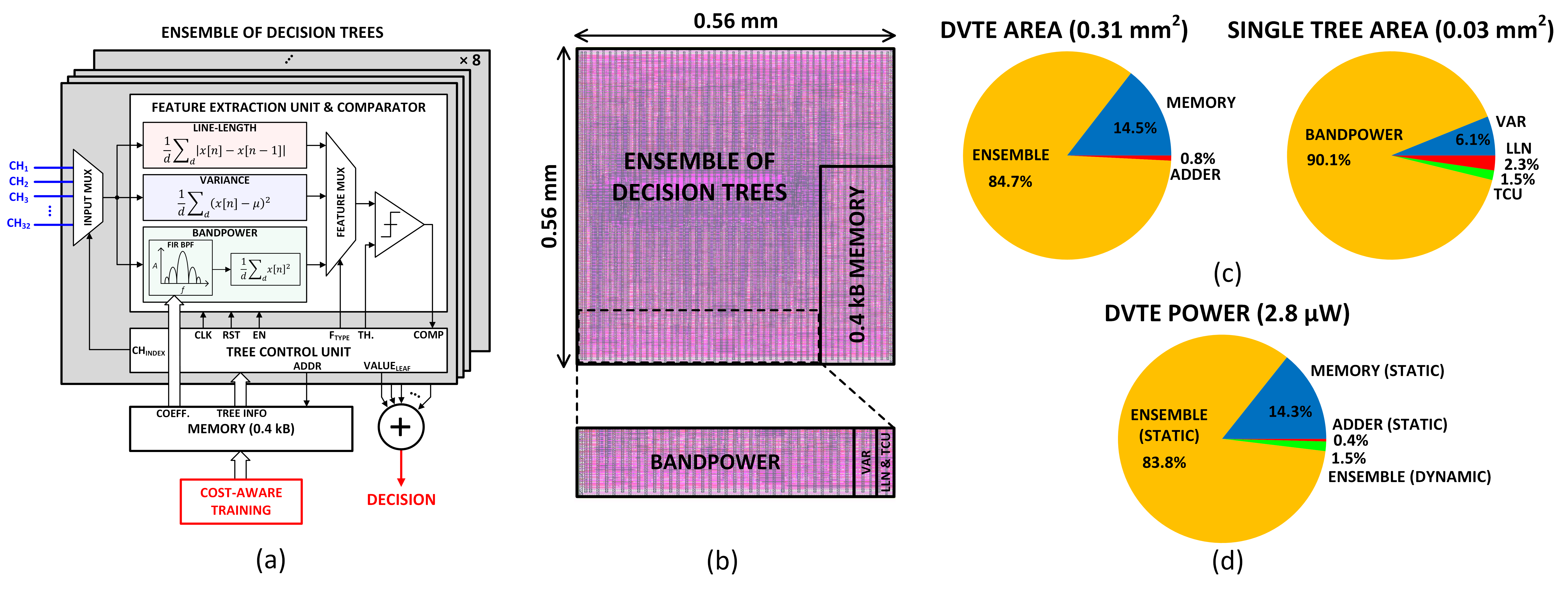}
	\vspace{-3mm}
	\caption{\textcolor{black}{Hardware implementation of the proposed DVTE classifier: (a) system architecture, (b) layout, (c) area breakdown of the DVTE processor and a single decision tree, and (d) system power breakdown.}}
	\label{fig_dvte_hd}
\end{figure*}

\vspace{-6mm}
\textcolor{black}{
\subsection{Hardware Implementation of DVTE Classifier}} \vspace{-1mm}
\textcolor{black}{We implemented the DVTE classifier in hardware to demonstrate the efficacy of the proposed cost-aware learning approach. Figure \ref{fig_dvte_hd}(a) presents the system architecture of the DVTE classifier, which supports 32-channel 500-S/s 10-bit input data. Each of the 8 decision trees consists of a feature extraction unit (FEU), a comparator, and a tree control unit (TCU). A 32-tap programmable FIR bandpass filter was implemented to extract the bandpower feature in a selected frequency band, and a single FIR coefficient memory was shared between 8 trees. The FEU extracts only one feature during each window, which allows us to clock- and data-gate unused feature blocks for dynamic power saving. The extracted feature is then compared to a threshold to decide the decision path in the tree. The TCU reads  the trained tree information (i.e., feature type, channel index, threshold, and leaf value) from memory and controls the FEU based on the current node information and comparison result. When a leaf node is reached, the tree sends out a leaf value and repeats the process starting from the root node. Leaf values from the 8 trees are summed to make a final decision. The proposed lightweight DVTE classifier utilizes a 0.4kB on-chip memory.}

\textcolor{black}{The DVTE classifier was implemented in a TSMC 65 nm 1P9M LP process. 
Figures \ref{fig_dvte_hd}(b) and (c) show the chip layout occupying 0.31 mm$^2$ and its area breakdown, respectively. Using Synopsys PrimeTime, the power consumption of the system was simulated at 2.8 $\mu$W under a 1.2-V supply. The parallel implementation of 8 trees allowed a low system clock (500 Hz). In addition, the use of high-Vt transistors saved both dynamic and static power consumption. The energy efficiency and E-A FoM of the DVTE classifier are 5.6 nJ/class. and 1.7 pJ$\cdot$mm$^2$/S, respectively, achieving $>$6.4$\times$ and $>$23.7$\times$ improvements over the state-of-the-art designs in Table. \ref{comparison1}.}

\textcolor{black}{In this advanced technology node with a low operating frequency and efficient clock- and data-gating, the static power consumption acts as the dominant source of power, as indicated in the breakdown of Fig. \ref{fig_dvte_hd}(d). Here, 83.8\% of  system power is consumed by  leakage currents in the ensemble. Therefore, power-gating of the unused feature extraction blocks can further improve the energy efficiency of the proposed cost-aware DVTE classifier. This is possible thanks to the on-demand feature extraction scheme of DVTE. To estimate the potential power savings, we performed post-layout simulations for each feature extraction block with power-gating header switches \cite{shi2006sleep}. The results showed that the static power consumption of each feature substantially reduced to 30 $p$W with the supply power gated. For the best trade-off case in Fig. \ref{f4} ($C=0.01$) with power-gating applied, the overall system power  is estimated to be 0.68 $\mu$W. It is our ongoing work to implement the power-gating technique reliably at the system level with a minimal area overhead, to potentially achieve sub-$\mu$W total power consumption.}


\section{Hardware-Algorithm Co-Design of \\Oblique Trees}
In the previous \textcolor{black}{section}, we proposed a novel tree ensemble, DVTE, and a cost-aware learning approach to improve  latency and power. However, tree ensembles may require a large number of axis-aligned DTs for non-trivial classification tasks \cite{zhu2019migraine,  o202026, yao2019enhanced}, resulting in a large model size and on-chip memory. Different from conventional trees that use axis-aligned decision boundaries, oblique trees calculate a weighted sum of multiple features and compare the result to a threshold \cite{zhu2020resot}. Thanks to their  powerful  split functions, oblique trees are capable of generating accurate predictions using a single tree with a reduced model size. Moreover, in our previous work, we built a  new class of oblique trees that are compatible with model compression techniques  to further reduce the  hardware complexity and memory needs \cite{zhu2020resot}. In this \textcolor{black}{section}, we present the hardware-algorithm co-design of oblique trees to simultaneously achieve low power consumption, low latency and small model size. 

We built oblique DTs using a probabilistic routing scheme \cite{zhu2020resot}, where the $i$-th internal node sends samples to a child according to the probabilistic distribution, as follows
\begin{equation}
P_i(\boldsymbol{{x}_n}) = softmax(\boldsymbol{{x}_n}^{\!\!\!\top}  \boldsymbol{\theta_i}),
\end{equation}
where $\boldsymbol{{x}_n}$ indicates  the feature vector and $\boldsymbol{\theta_i}$ is the trainable weight vector of the same shape as $\boldsymbol{{x}_n}$. The $softmax$ function normalizes the output space into a probability distribution within (0,1) interval. Here, $\boldsymbol{{x}_n}$ visits the left child with a probability of $P_i(\boldsymbol{{x}_n})$ and the right child with $1-P_i(\boldsymbol{{x}_n})$. In the probabilistic routing scheme, samples arrive at multiple leaf nodes with different probabilities and the final prediction is given by
\begin{equation}
\hat{y}_n = \sum_l P_l(\boldsymbol{{x}_n}) \omega_l,
\end{equation}
where $P_l(\boldsymbol{{x}_n})$ indicates the probability of sample $\boldsymbol{{x}_n}$ visiting the leaf node $l$ and $\omega_l$ is the constant leaf predictor. For classification tasks, we measured the cross-entropy loss $\sum_n L(\hat{y}_n, y_n)$ using the groundtruth ($y_n$) and the prediction of the oblique tree $\hat{y}_n$.

\vspace{-2mm}
\subsection{Model Compression and Cost-Aware Learning} \vspace{-1mm}
Various compression techniques have been applied to DNNs, including fixed-point quantization \cite{lin2016fixed}, weight pruning and sharing \cite{han2015deep}. Interestingly, within the probabilistic training scheme, oblique trees are compatible with gradient descent-based optimization, similar to the training of a neural network. Therefore, we propose to combine oblique trees with DNN-based compression techniques to reduce model size and hardware cost.
We trained the oblique tree by minimizing the loss $\sum_n L(\hat{y}_n, y_n)$ on training data. During the training process, we applied weight pruning to slim the oblique tree and weight sharing to further reduce model size. Specifically, we used a simple neural network with  input and output layers  to represent the oblique decision functions in the internal nodes. Weight pruning/sharing were applied to  2-layer NNs for creating sparse connections and reducing the model size. We pruned the oblique tree by iteratively setting small values to zero and retraining the remaining parameters. For weight sharing, we uniformly clustered the weights into $k$ shared values, requiring only $\ceil*{log_2 k}$ bits to store the index. It should be noted that oblique trees are compatible with the aforementioned cost-aware learning framework, by simply replacing the loss function in Eq. \ref{eq1} with oblique tree training loss. In cost-aware learning, oblique trees assign smaller weights to costly features so that they hardly survive the  pruning process. 

\begin{figure}[t]
  \centering
  \includegraphics[width=1\columnwidth]{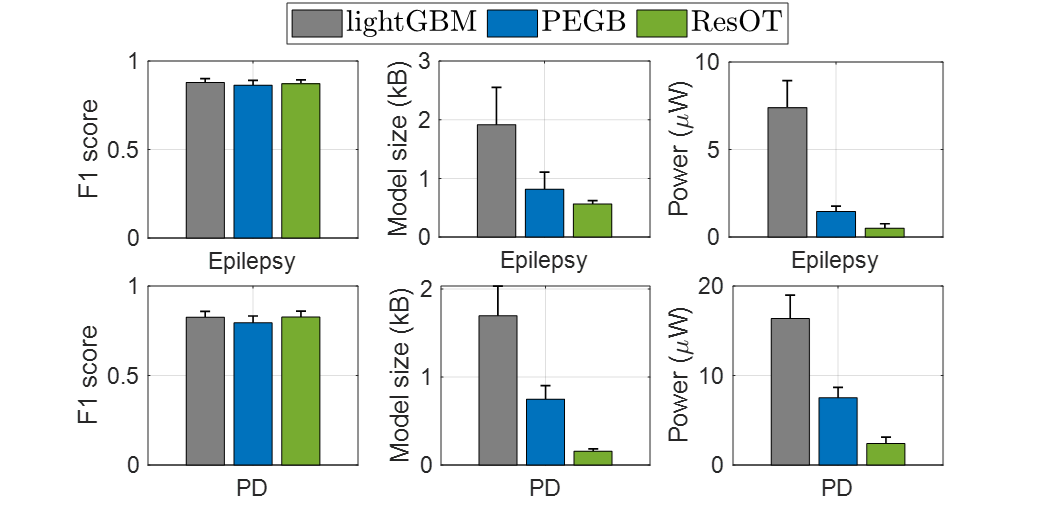}
  \vspace{-8mm}
  \caption{Comparison of ResOT and axis-aligned tree ensemble on seizure and tremor detection tasks. 
  The conventional gradient boosted ensemble (lightGBM \cite{ke2017lightgbm}) and  gradient boosting with power-efficient training (PEGB) were included. For PEGB, We used fixed-point  thresholds  and leaf weights, in contrast to floating points in lightGBM  \cite{zhu2019cost}. Cross-subject standard errors are shown by error bars.
  }\vspace{-4mm}
  \label{fig5}
\end{figure}

We compared the hardware efficiency of oblique trees against axis-aligned tree ensembles. Specifically, we built resource-efficient oblique trees (ResOT) \cite{zhan2019resource} by combining cost-aware learning with  model compression. We used the conventional lightGBM ensemble \cite{ke2017lightgbm} and  gradient boosting with power-efficient training (PEGB \cite{zhu2019cost}) as baseline. 
In addition to seizure detection, we tested our model on LFPs  recorded from 12 PD patients 
via DBS leads (3-channel, 2048~Hz  sample rate, 16 recordings) to detect the tremor onset \cite{yao2020improved}. 
For both tasks,  a single ResOT was built (max depth: 4) with 16  shared weights (4 bits). Hyperparameters of oblique trees including the number of parameters post pruning and the regularization coefficient were optimized for each patient.
\textcolor{black}{We used  5-fold chronological cross-validation to measure the F1 score, and  leave-one-out  for epilepsy patients with $<$5 seizures. 
Cross-validation has been widely used in previous studies \cite{o2018nurip, shoeb2010application, shoaran2018energy}. It allows testing on multiple train-test splits to fairly assess the model performance  on unseen data. Compared to the hold-out method \cite{altaf201516}, cross-validation is less dependent on a specific train-test split and could provide a reliable measure of performance for patients with few seizure events.
We employed a block-wise data splitting method, where each block includes a complete seizure event and its neighbouring non-seizure period, to avoid information leakage during training \cite{shoaran2018energy}. Cross-validation was performed on pre-recorded data to estimate the model performance and optimize the hyperparameters. In a clinical setting, the final set of parameters (i.e., feature index, threshold, leaf weights, and feature weights for oblique trees) will be trained using the entire pre-recorded data of each patient and loaded to the chip to predict future seizure events. We estimated the memory requirements of various models using the size of the trainable weight matrix. Compressed sparse column and delta encoding were used to store the sparse matrix after weight pruning. For power comparison, we considered the power consumption for extracting features along the decision path during inference (Table. \ref{tab1}).} In our simulations, ResOT achieved an average saving of 7.0$\times$ in model size and 10.7$\times$ in power cost compared to lightGBM, as shown in Fig.~\ref{fig5}.  It  also outperformed the hardware-efficient ensemble (PEGB) (3.1$\times$ in model size and  3.0$\times$ in power cost). 

\begin{figure}[b]
  \centering
  \includegraphics[width=1\columnwidth]{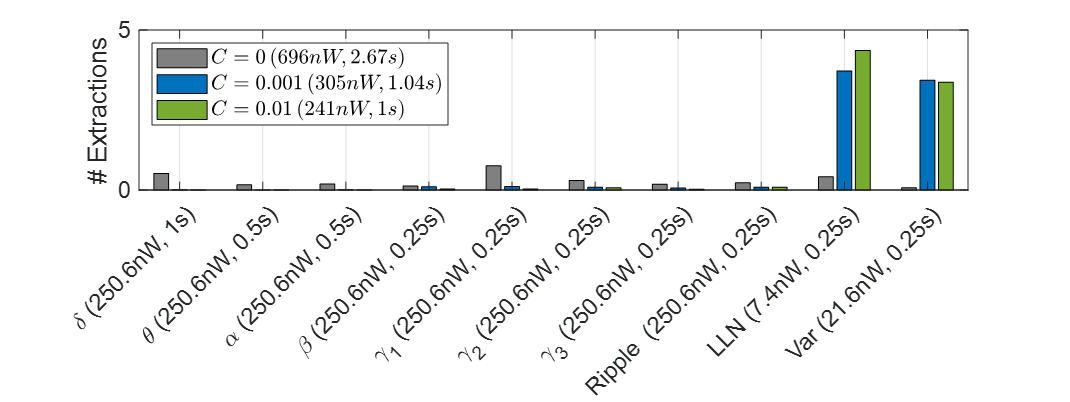}
  \vspace{-6mm}
  \caption{The number of extracted features with different regularization coefficients in ResOT. With greater power- and latency-aware regularization terms, oblique trees  prioritize low-power and low-latency features.}\vspace{-3mm}
  \label{f10}
\end{figure}

The oblique node evaluation time is set by the longest feature computed in that node. Here, we pruned the oblique tree to use a maximum of 8  features per internal node to fairly compare it against DVTE. The hardware cost  in an oblique tree (e.g., power, latency) can also benefit from the introduced cost-aware learning scheme, by including a cost regularization in the oblique tree objective. Figure \ref{f10} plots the distribution of  features in ResOT on  seizure detection task. The latency was reduced from 2.67s to $\sim$1s via cost-aware training, while the power cost was reduced from \textcolor{black}{696nW} to \textcolor{black}{241nW}.

\vspace{-2mm}
\subsection{Parallel Node Evaluation for Latency Reduction} 
\begin{figure}[t!]
  \centering
  \includegraphics[width=1\columnwidth]{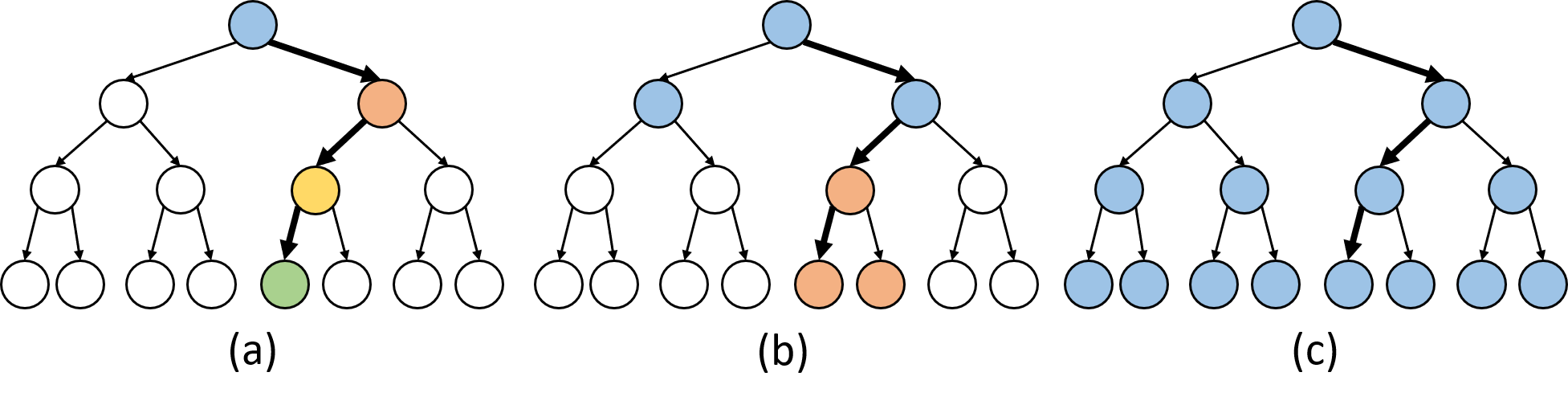}
  \vspace{-7mm}
  \caption{Parallel node evaluation scheme. (a) One internal node is evaluated per window. (b)  Two layers (maximum 3 nodes) are  concurrently evaluated per window. (c) All nodes are evaluated in parallel. The evaluated nodes are shown in color and  bold lines  represent the decision path. 
  Node colors represent successive windows.}
  \label{f7}
\end{figure}

\vspace{-1mm}
Previous work on oblique trees employed a single-path inference scheme by visiting the most probable path, which suffers from a latency proportional to the length of the decision path \cite{zhu2020resot}. An alternative is to evaluate multiple nodes in parallel to reduce latency,
as shown in Fig. \ref{f7}. The single-path inference scheme is presented in Fig. \ref{f7}(a), where 4 internal nodes are evaluated using  consecutive windows. In Fig. \ref{f7}(b), we evaluate two layers of the tree (3 nodes) per  window, requiring  6  node evaluations in total. Finally, Fig.~\ref{f7}(c) evaluates all 15 nodes in parallel. 

\begin{figure}[t!]
  \centering
  \includegraphics[width=0.8\columnwidth]{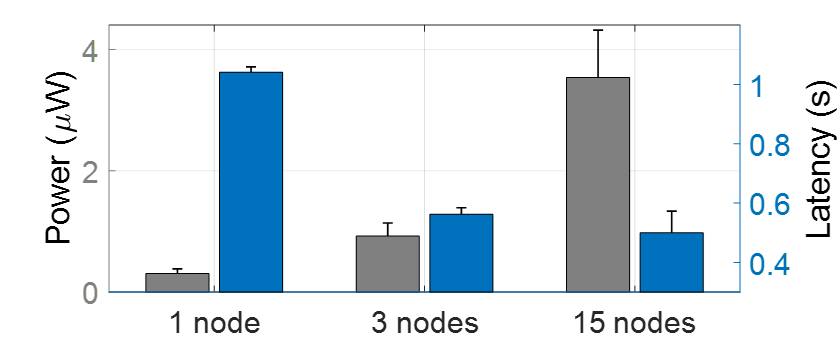}
  \vspace{-1mm}
  \caption{The power-latency trade-off with parallel node evaluation. With more nodes evaluated in parallel, latency is reduced at the cost of increased power consumption. Experiments were conducted with ResOT on  epilepsy task.}\vspace{-3mm}
  \label{f8}
\end{figure}

To demonstrate the trade-off between power consumption and latency, we built an oblique tree on seizure detection task. 
Different numbers of nodes were evaluated in parallel and  the corresponding hardware cost is reported in  Fig. \ref{f8}. Single-path inference (Fig. \ref{f7}(a)) obtained the lowest power  and highest latency (power cost = \textcolor{black}{305$nW$}, latency = 1.04s). On the other hand, concurrently evaluating all  nodes in parallel  reduced the latency by 2.1$\times$ but increased the power by 11.6$\times$ (Fig. \ref{f7}(c)). The case of  3 nodes (Fig. \ref{f7}(b)) achieved a better trade-off between power and latency, leading to 1.9$\times$ reduction in latency and  3$\times$ increase in power cost. This scheme can potentially be useful in latency-constrained applications.

\begin{figure*}[t]
  \centering
  \includegraphics[width=1.85\columnwidth]{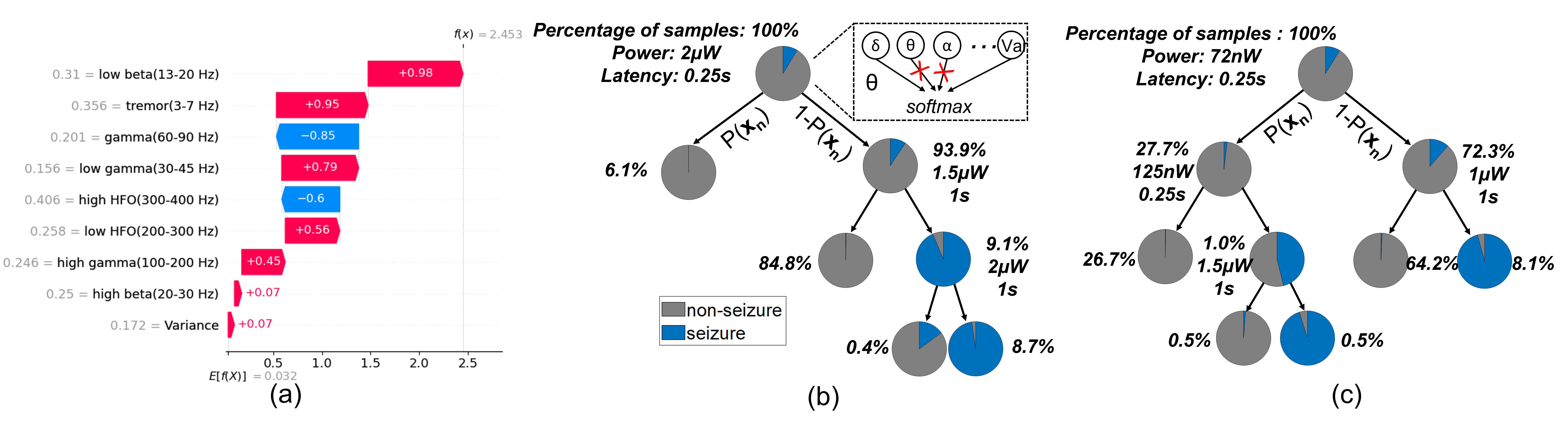}
  \vspace{-3mm}
  \caption{(a) Interpretation of the tremor detection process using a tree ensemble and shapley additive explanations. Features  plotted in red predicted an increased risk of tremor, while those in blue were associated with a low tremor risk. (b) Visualization of the seizure detection process in an oblique tree in one arbitrary patient. 
  We show the percentage of samples visiting each node, and the required power and latency to evaluate each internal node. There are multiple “short paths” which allow  dynamic early exiting. (c) Visualization of a cost-aware oblique tree, showing a significant reduction in the power cost of the root node. }\vspace{-3mm}
  \label{f9}
\end{figure*}

\vspace{-2mm}
\subsection{Interpretable DTs for Neural Prostheses}
Closed-loop stimulation is a safety-critical application, favoring an interpretable decision process. Another distinct advantage of tree-based models is their interpretability, in contrast to most classical machine learning and deep learning methods that lack transparency and interpretability. This is critical to understanding a specific therapeutic strategy for a particular neurological symptom or behavior. We can simply visualize the decision process of DTs and the informative biomarkers used in making predictions. 
Therefore, tree-based models are widely used in clinical applications that require high interpretability  \cite{artzi2020prediction ,lundberg2018explainable}. 

For example, Fig.\ref{f9}(a) shows the contributions from time- and spectral-domain features in tremor detection task, using shapley additive explanations \cite{NIPS2017_7062}. The feature values  at the visualized window are shown on the left, and the  red/blue colors  represent features that indicate a high/low risk of tremor, respectively. The power over low beta and tremor bands are the most predictive features. The model predicts a tremor state according to the weighted contribution of all features.  

Figure \ref{f9}(b)  visualizes the decision process of an oblique  tree on seizure detection task. We used pie charts at  internal and leaf nodes to represent the class distribution. Both seizure and non-seizure samples are mixed at the internal nodes, while each leaf node is dominated by either seizure or non-seizure samples. The decision process follows an explainable rule list structure, with the left branch leading samples directly to a leaf node. 
The percentage of samples that travel through a node (internal or leaf) is shown next to that node. 
We also show the approximate power and latency to process each internal node. For comparison, Fig. \ref{f9}(c) shows the decision process of a cost-aware oblique tree trained on the same patient. As shown in this figure, the power cost to evaluate the internal nodes is significantly reduced in the cost-aware approach. Particularly, the most notable reduction in power (i.e., node complexity) is  observed at the  root node, as it is the most frequently visited node in the tree. \textcolor{black}{Moreover, 
the overall latency along the root-leaf  path in Fig. \ref{f9}(c) is shorter than that of Fig. \ref{f9}(b), indicating a reduction of processing latency.}

\vspace{-2mm}
\section{Conclusion}\vspace{-1mm}
In this paper, we reviewed the latest developments in closed-loop neural interface design, with a particular focus on system-on-chips that integrate machine learning for symptom detection. The current commercial and research-based closed-loop devices, advances in electrode and circuit design, and clinical applications were discussed.
We reviewed various hardware approaches used to implement machine learning  on  neural prostheses, design trade-offs and hardware/performance comparisons. 
We further proposed a novel tree-based neural decoder, Depth-Variant Tree Ensemble, to reduce  latency in neurological symptom detection. A cost-aware learning approach was applied to DVTE to further reduce power and latency. We also integrated various techniques, including cost-aware learning and model compression, to construct resource-efficient oblique trees. Testing on epileptic seizure and PD tremor detection tasks, the proposed model improved both power and latency, and reduced the memory requirement, while maintaining a high  performance. We also discussed 
the interpretability of tree-based models, as an essential component for next-generation intelligent neural prostheses.

\vspace{-2mm}
\section*{Acknowledgment}
This work was partially supported by the National Institute of Mental Health Grant R01-MH-123634.


\ifCLASSOPTIONcaptionsoff
  \newpage
\fi



%

\vspace{-1mm}
\bibliographystyle{IEEEtran}
\bibliography{cit}

\end{document}